\documentclass[aps,pre,preprint, superscriptaddress]{revtex4-1}
\usepackage{latexsym}
\usepackage{natbib}
\usepackage{amsmath}
\usepackage{amsfonts}
\usepackage{amssymb}
\usepackage[utf8]{inputenc}
\usepackage[T1]{fontenc}
\usepackage[dvips]{color,graphicx}   %para insertar imagenes
\usepackage{verbatim}   % useful for program listings
\usepackage{t1enc}
\usepackage{anysize}
\usepackage{tocbibind}
\usepackage[percent]{overpic}
\usepackage{float} 
\usepackage{appendix} %sirve para poner apéndice
\usepackage{booktabs} 
\usepackage{natbib}
\usepackage{array}
\usepackage{xcolor}
\usepackage[normalem]{ulem}

\begin{document}
\title{Explosive epidemic transitions induced by quarantine fatigue}
\author{L.  D. Valdez}\email{} 
\affiliation{Departamento de F\'isica, FCEyN, Universidad Nacional de Mar del Plata, Mar del Plata 7600, Argentina.}
\affiliation{Instituto de Investigaciones F\'isicas de Mar del Plata (IFIMAR), CONICET, Mar del Plata 7600, Argentina.}
  \date{\today}

\begin{abstract}
Quarantine measures are one of the first lines of defense against the spread of infectious diseases. However, maintaining these measures over extended periods can be challenging due to a phenomenon known as quarantine fatigue. In this paper, we investigate the impact of quarantine fatigue on the spread of infectious diseases by using an epidemic model on random networks with cliques. In our model, susceptible individuals can be quarantined up to $n$ times, after which they stop complying with quarantine orders due to fatigue. Our results show that quarantine fatigue may induce a regime in which increasing the probability of detecting and isolating infected individuals (along with their close contacts) could subsequently increase the expected number of cases at the end of an outbreak. Moreover, we observe that quarantine fatigue can trigger an abrupt phase transition at the critical reproduction number $R_0=1$. Finally, we explore a scenario where a non-negligible number of individuals are infected at the beginning of an epidemic, and our results show that, depending on the value of $n$, an abrupt transition between a controlled epidemic and a large epidemic event can occur for $R_0<1$. 

\end{abstract}

\maketitle

\section{Introduction}
During the early stages of the COVID-19 pandemic,  numerous non-pharmaceutical measures, including quarantine orders and mass testing, were implemented across multiple communities and regions at an unprecedented scale in order to reduce the daily number of new cases~\cite{kwok2020comparing,robin2022local}. However, after applying these measures for several consecutive months, it was clear that many individuals became less compliant with quarantine guidelines due to psychological and economic factors~\cite{sood2021pandemic,zhao2020quarantine}. This concerning phenomenon known as quarantine fatigue or pandemic fatigue could adversely affect the course of epidemics~\cite{meacci2021pandemic}. Therefore, understanding the effects of this phenomenon becomes essential for the development of effective public health interventions.

Throughout history, quarantine has played a key role in the control of re-emerging and newly emerging diseases~\cite{cliff2009emergence}. According to the CDC, quarantine is defined as a restriction of the movement of individuals who have been exposed to a contagious disease to see if they become sick~\cite{CDCdefinitions}. Recent examples in which this strategy has been implemented include SARS, Ebola, and COVID-19~\cite{conti2020historical,gensini2004concept,kalra2014emergence}. Another term closely related to quarantine is "isolation," which involves the separation of infected {\bf} individuals from those not infected~\cite{CDCdefinitions}. In this work, we will use the terms quarantine and isolation interchangeably to refer to the separation of an individual (whether or not infected) from the general population to prevent the transmission of a contagious disease.

Numerous mathematical and computational models have been developed to explore the effectiveness of quarantine measures~\cite{nussbaumer2020quarantine}. For instance, several studies have shown that a significant percentage of cases and deaths could potentially be prevented by isolating individuals who have been in contact with confirmed or suspected cases~\cite{nussbaumer2020quarantine,ferguson2020report,koo2020interventions,semenova2020epidemiological}.  These models also showed that quarantine measures are able to reduce the basic reproduction number by more than half~\cite{nussbaumer2020quarantine,kucharski2020effectiveness,madubueze2020controlling}. On the other hand, Aleta et al.~\cite{aleta2020modelling} found that a mixed approach involving contact tracing and quarantine measures might successfully prevent a second wave of infections.

During the COVID-19 crisis, although quarantine measures and other non-pharmaceutical interventions have contributed to containing the spread of this virus, concerns have arisen about the viability of these measures over prolonged periods. For example, a survey conducted among 516 adults in Turkey showed that one-third were taking fewer COVID-19-related precautions in November 2020 compared to the beginning of the pandemic~\cite{haktanir2022we}. Moreover, Taylor et al.~\cite{taylor2022develops} found that the phenomenon of quarantine fatigue was associated with pessimism and apathy (predominantly among younger individuals). Similarly, Williams et al.~\cite{williams2021public}  noted that non-adherence to pandemic protection measures could also be caused by other reasons including a lack of trust in government, learned helplessness, alert fatigue, and a reduced perception of risk.

To better understand the impact of fatigue on the spread of infectious diseases, a few mathematical models have been developed in recent years~\cite{de2021effect,meacci2021pandemic,collinson2013modelling,gualtieri2021sars}. For example, de Meijere et al.~\cite{de2021effect} studied an epidemic model in complex networks where quarantined individuals can prematurely end their isolation before fully recovering. Their work showed that populations are more vulnerable to epidemics when individuals with more social contacts tend to be less compliant with quarantine rules. In another work, Meacci and Primicerio~\cite{meacci2021pandemic} developed a mean-field model in which the infection rate $\beta$ is controlled by two mechanisms: pandemic fatigue (which increases $\beta$) and the number of deceased individuals (which decreases $\beta$). On the other hand, Collison~\cite{collinson2013modelling} proposed an epidemic model to investigate the impact of the media on disease spread. Specifically, they studied the case in which individuals, guided by media reports of the epidemic, opt for isolation and vaccination. However, the impact of these reports progressively diminishes over time because the public becomes less emotionally sensitive to negative or aversive stimuli after repeated exposure~\cite{collinson2013modelling}. In their model, they observed that desensitization to the media slightly increases the number of infected individuals and the duration of the epidemic.

Recent research on the role of fatigue in the spread of diseases was primarily focused on variations of the standard susceptible-infected-susceptible (SIS) and susceptible-infected-recovered (SIR) models. Typically, these two models exhibit a continuous phase transition when the basic reproduction number $R_0$ is equal to 1~\cite{pastor2015epidemic}. However, it has been observed that certain epidemic models, particularly those where the effectiveness of non-pharmaceutical interventions declines with time~\cite{bottcher2015disease,borner2022explosive,valdez2023epidemic} can exhibit a discontinuous transition. This may occur, for instance, because there are limitations in the resources available to prevent disease transmission~\cite{lamata2023collapse,bottcher2015disease,borner2022explosive,chen2019nontrivial,scarselli2021discontinuous,di2018multiple}. For example, B\"orner et al.~\cite{borner2022explosive} proposed a mean-field model with a capacity-limited intervention and it was observed that their model can exhibit an explosive epidemic transition. On the other hand, Valdez et al.~\cite{valdez2023epidemic} studied a prompt quarantine strategy in networks with cliques, based on a previous work published in Ref.~\cite{hasegawa2017efficiency}. They found that around $R_0=1$, the system can undergo a continuous or discontinuous phase transition depending on the model parameters. In addition, it was observed that their model is sensitive to initial conditions and epidemic outbreaks can occur even when $R_0<1$, a phenomenon termed "backward bifurcation" in the language of non-linear systems~\cite{gumel2012causes,valdez2023epidemic}.

In this paper, we extend our model presented in Ref.~\cite{valdez2023epidemic} to investigate the influence of quarantine fatigue on random networks with cliques. Specifically, we incorporate quarantine fatigue into our model by assuming that susceptible individuals will comply with quarantine orders up to $n$ times, after which they will no longer obey due to fatigue.   For this model, we study two variants, namely Strategy A and Strategy B, which we will explain in Sec.~\ref{Sec.model}.  Our results reveal that a phase transition occurs at the critical reproduction number $R_0=1$. Moreover, we find that there exists a regime in which this transition is continuous in the absence of fatigue, while for small values of $n$, the transition becomes discontinuous.

Finally, we consider a scenario where there is a macroscopic fraction of infected individuals at the beginning of disease spread, and our numerical results show that the transition point between a controlled epidemic and an explosive epidemic depends on the fatigue threshold $n$.

This paper is organized as follows. In Sec.~\ref{Sec.model}, we present our discrete-time SIRQ (susceptible-infected-recovered-quarantined) model with fatigue. Secs.~\ref{sec.ResMicro}-\ref{sec.ResMacro} present our numerical results for microscopic and macroscopic initial conditions. In Sec.~\ref{Sec.Conclu}, we display our conclusions. Additionally, we enhance this work with numerous appendices that extend our findings.

\section{Model}\label{Sec.model}
Over the past twenty years, researchers have investigated how different topological properties of social networks affect the processes that occur on top of these networks~\cite{boccaletti2006complex,battiston2020networks,newman2018networks}. For instance, social networks often exhibit heterogeneity (meaning that some individuals, known as hubs, have a disproportionately large number of connections, while most people are poorly connected to the network) and research has shown that this property can facilitate the spread of contagious diseases~\cite{pastor2015epidemic,newman2018networks,meyers2007contact}. Additionally, social networks frequently contain densely connected groups, or cliques, which have been found to play a significant role in the dynamics of disease transmission and outbreak control~\cite{st2021social,rizi2023unreasonable,ma2013effective}.

\begin{figure}[H]
\begin{center}
\begin{overpic}[scale=0.9]{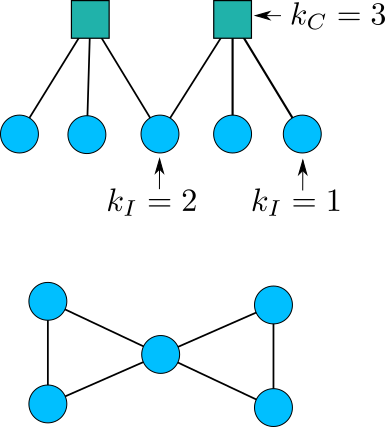}
  \put(0,95){(a)}
  \put(0,25){(b)}
\end{overpic}
\end{center}
\caption{Panel a: Illustration of a bipartite network. Squares represent cliques, while circles represent individuals. Panel b: One-mode projection of the bipartite network (shown in Panel a) onto the set of individuals.}\label{fig.Bip}
\end{figure}

In our work,  we will focus on networks with cliques, which can be represented as bipartite networks~\cite{newman2018networks}, as shown in Fig.~\ref{fig.Bip}. Here, a bipartite network consists of two distinct sets of nodes: one set represents the cliques, and the other one represents the individuals. Further details on the network parameters are given in Table~\ref{tab:Def}, while the methodology for constructing these networks is provided in Appendix~\ref{App.Bip}. 

\begin{table}[H]
\caption{Parameters and variables} \renewcommand{\arraystretch}{1}
%esto es para aumentar la separaci\'on entre filas
\begin{tabular}{lp{0.11cm}}  %% el @{} es para agregar espacio
\toprule[0.05cm]
\multicolumn{1}{c}{}&\multicolumn{1}{l}{Definition}\\\cmidrule{1-2}
\multicolumn{1}{c}{$N_I$}&\multicolumn{1}{l}{Number of individuals}\\
\multicolumn{1}{c}{$N_C$}&\multicolumn{1}{l}{Number of cliques}\\
\multicolumn{1}{c}{$P(k_I)$}&\multicolumn{1}{l}{Probability distribution that an individual belongs to $k_I$ cliques}\\
\multicolumn{1}{c}{$P(k_C)$}&\multicolumn{1}{l}{Probability distribution that a clique contains $k_C$ individuals}\\
\multicolumn{1}{c}{$\beta$}&\multicolumn{1}{l}{Infection probability.}\\
\multicolumn{1}{c}{$f$}&\multicolumn{1}{l}{Detection probability}\\
\multicolumn{1}{c}{$t_r$}&\multicolumn{1}{l}{Recovery time.}\\
\multicolumn{1}{c}{$t_b$}&\multicolumn{1}{l}{Quarantine or isolation time}\\
\multicolumn{1}{c}{$n$}&\multicolumn{1}{l}{Maximum number of times a susceptible individual will obey a quarantine order}\\
\multicolumn{1}{c}{$R$}&\multicolumn{1}{p{12cm}}{Fraction of recovered people}\\
\multicolumn{1}{c}{$I_0$}&\multicolumn{1}{p{12cm}}{Fraction of infected people at the beginning of the disease spread}\\
\bottomrule[0.05cm]
\end{tabular}
\label{tab:Def}
\end{table}

On these networks, we will investigate two variants of our SIRQ model (with quarantine fatigue), specifically Strategy A and Strategy B, which we will explain below.

For Strategy A, at each time step ($t \to t+1$), the following sequence is performed:
\begin{itemize}
\item Step 1) Detection and quarantine: Each infected individual is detected and isolated with probability $f$.
\item Step 2) Contact tracing and quarantine: The neighbors of the isolated individuals from the previous step are also placed in quarantine.
\item Step 3) Infection: Infected individuals not detected or isolated in the previous steps have a chance of transmitting the disease to their susceptible neighbors with probability $\beta$.
\item Step 4) Recovery: Individuals recover after being infected for a period of $t_r$ time steps. Additionally, individuals who have been placed under quarantine in steps 1 and 2 also shift to the recovered state after $t_r$ time steps.
\item Step 5) Return and Quarantine Compliance: Susceptible individuals who were quarantined in step 2 will return to the network after $t_b$ time steps. These individuals comply with quarantine orders up to $n$ times. Once this limit is reached, they stop complying and refuse further testing. However, infected individuals who were isolated in Steps 1 or 2 will not rejoin the network until they have fully recovered.
\end{itemize}

Note that under Strategy A, individuals are detected and isolated before they can infect others. On the other hand, the specific case where $t_b=\infty$ represents a scenario explored in depth in~\cite{valdez2023epidemic}, where isolated individuals remain permanently disconnected from the network.

Now, turning to our alternative model, Strategy B follows almost the same sequence of steps described for Strategy A, with only one key difference: the infection step is placed as the first step during the transition from time $t$ to $t+1$. Consequently, under Strategy B, infected individuals have the opportunity to transmit the infection before being detected. 

The source codes (written in FORTRAN) for Strategy A and B are available on GitHub~\cite{gith01}.

Regarding the initial conditions at $t=0$, we will consider two cases:
\begin{itemize}
    \item a microscopic initial condition, where only one person is infected (referred to as "patient zero" or "index case"), and the rest of the population is susceptible.
    \item a macroscopic initial condition, where a fraction $I_0$ of individuals is infected, while the rest of the population is susceptible.
\end{itemize}

Once the disease stops spreading (because the number of infected people goes to zero), we measure the fraction of recovered people. If a significant portion of the population is recovered at the final stage, we classify this outbreak as an epidemic~\cite{lagorio2009effects}. Conversely, if only a small number $s$ of people is recovered at the final stage (i.e. $s \ll N_I$),  this outbreak is classified as a "small outbreak."

In the subsequent sections, we will present our results for both Strategy A and Strategy B for $t_r=1$, and $t_b=1$. In the Supplemental Material, we include more results for $t_r=5$, and $t_b=1$ and $t_b=5$.

\section{Results for a microscopic initial condition}\label{sec.ResMicro}
In this section, we present the numerical results of our SIRQ model (starting from a single infected individual at $t=0$) on networks with cliques. Given the diversity of network topologies to explore, we will focus here only on the simplest case, which corresponds to networks where every individual belongs to the same number $k_I$ of cliques, and all cliques have an equal number $k_C$ of individuals. We refer to these networks as "random regular (RR) networks with cliques." However, in the Supplemental Material, we expand our analysis to include results for networks with other topologies. 

\begin{figure}[H]
\begin{center}
\begin{overpic}[scale=0.40]{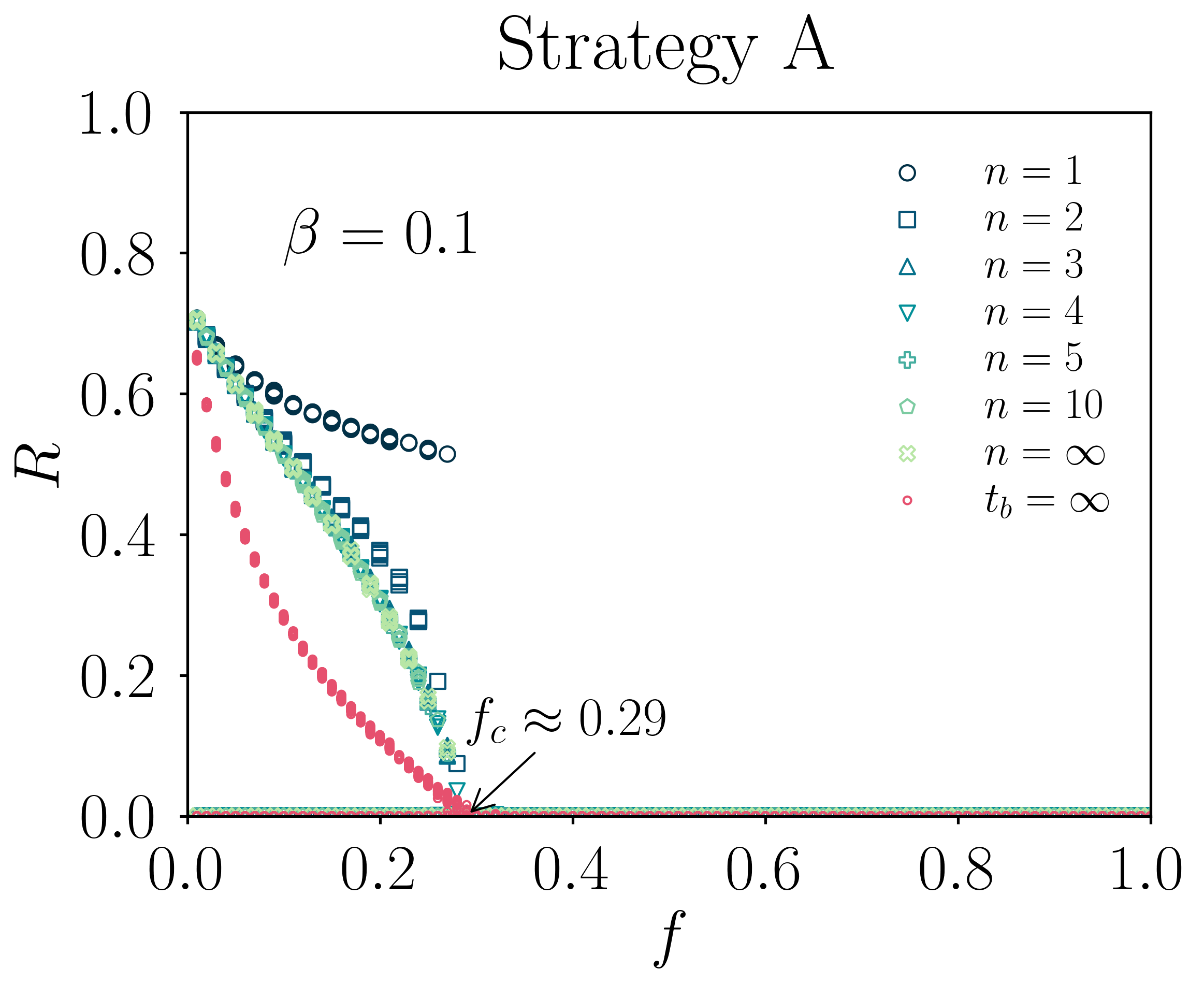}
  \put(85,18){(a)}
\end{overpic}
\begin{overpic}[scale=0.40]{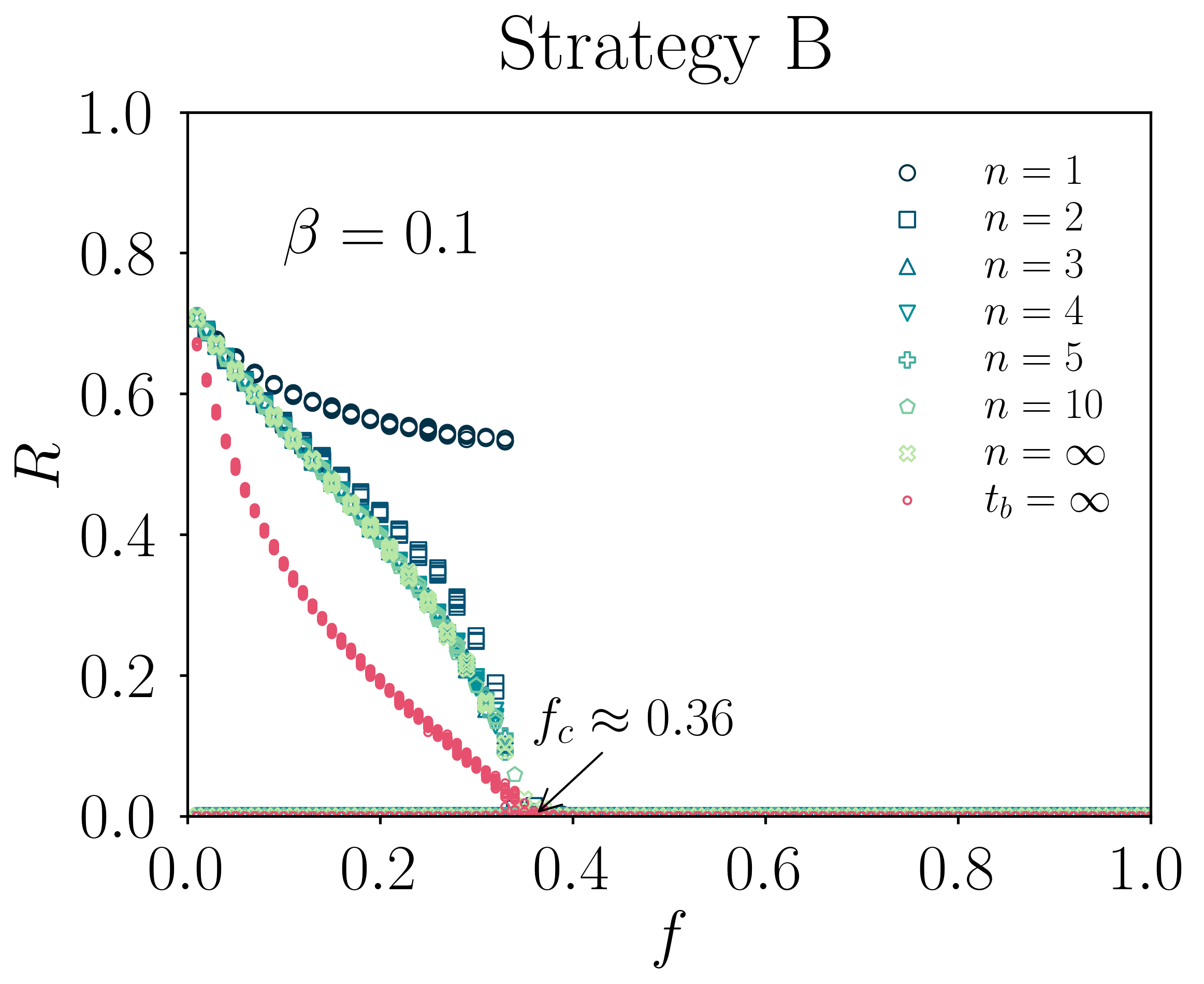}
  \put(85,18){(b)}
\end{overpic}
\begin{overpic}[scale=0.40]{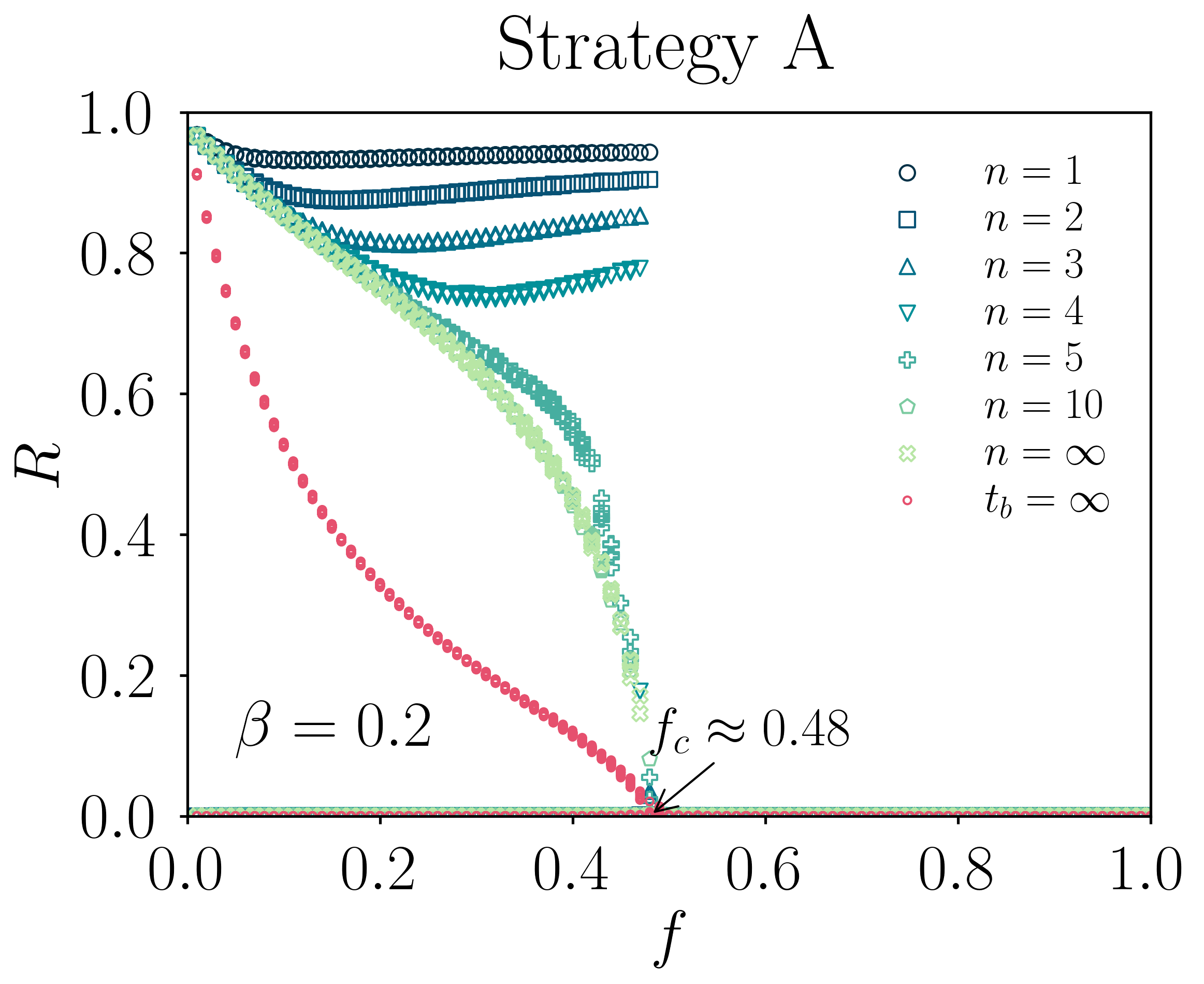}
  \put(85,18){(c)}
\end{overpic}
\begin{overpic}[scale=0.40]{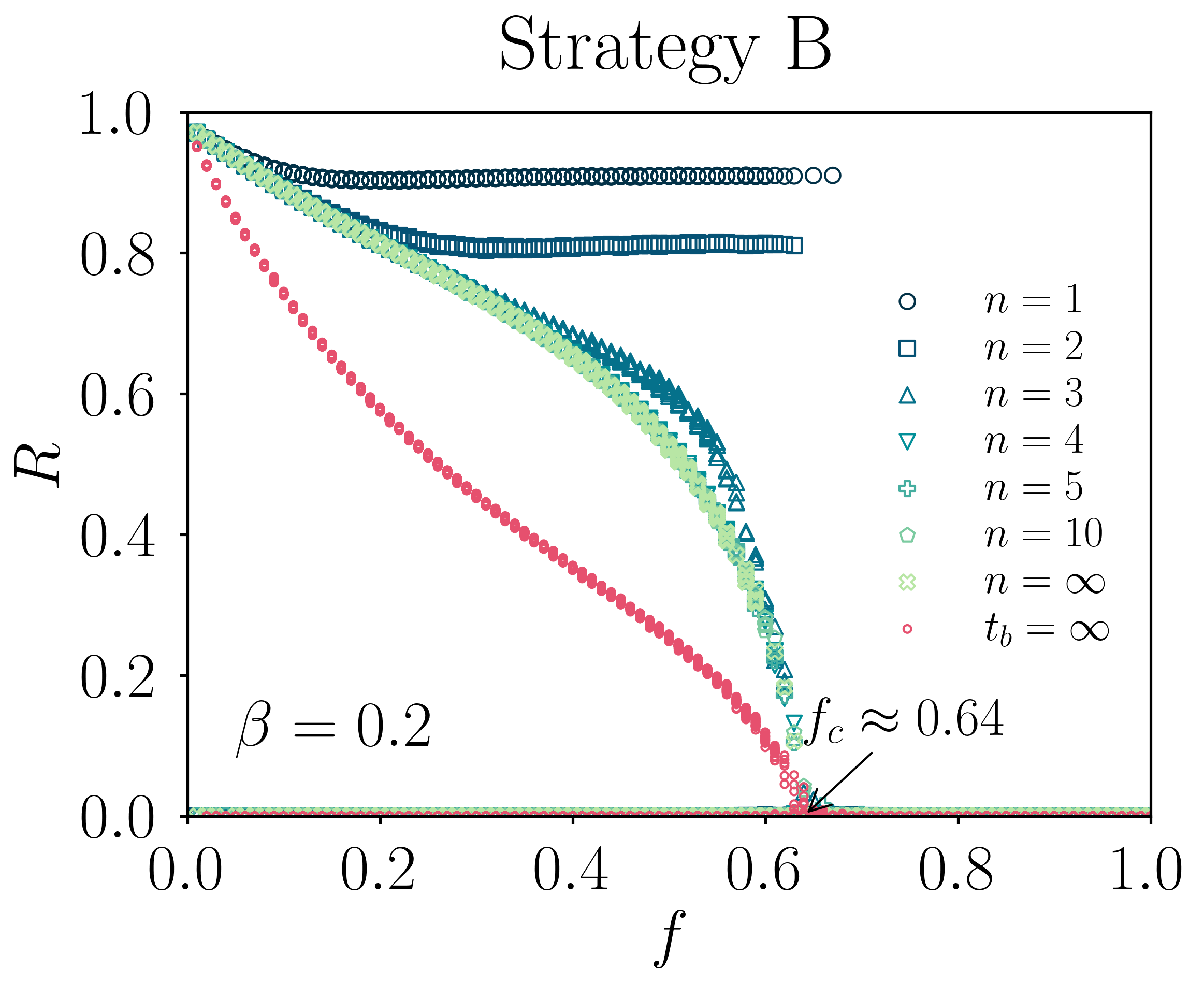}
  \put(85,18){(d)}
\end{overpic}
\vspace{-1.1cm}
\end{center}
\caption{Scatter-plot of $R$, as a function of
  $f$ for Strategy A [panels (a) and (c)] and Strategy B [panels (b) and (d)], for $t_r=1$, $t_b=1$ and considering various values of the probability of infection ($\beta$) and fatigue thresholds ($n$). Additionally, we also include the case $t_b=\infty$, where isolated individuals never rejoin the network. All simulations were performed on RR networks with $k_C=7$, $k_I=3$ and $N_I=10^6$. Results were obtained from $200$ stochastic realizations. The arrows indicate the threshold $f_c$ estimated from: 1) Eqs.~(\ref{eq.StA.R0})-(\ref{eq.e2StratA}) for Strategy A, and 2) Eqs.~(\ref{eq.StB.R0})-(\ref{eq.e2StratB}) for Strategy B (see Sec.~\ref{Sec.R0Calc}).}\label{fig.scatt}
\end{figure}

In Figs.~\ref{fig.scatt}a-d, we show the fraction of recovered individuals at the final stage as a function of $f$ for a fixed value of $\beta$ and various values of the fatigue threshold, $n$. In addition, we include the results for our SIRQ model with $t_b=\infty$ which represents the scenario explored in Ref.~\cite{valdez2023epidemic} in which isolated individuals never rejoin the network. 

Several observations can be made from these figures. On one hand, we note that the transition point, denoted as $f_c$ (above which the population is in a disease-free phase) does not depend on the fatigue threshold ($n$) and its value coincides with the threshold observed for the case of $t_b=\infty$. On the other hand, we can see that for both strategies and large $n$, the transition at $f=f_c$ is continuous and the fraction of recovered people decreases monotonically with increasing $f$.  

However, as the value of $n$ decreases, the transition becomes discontinuous at $f_c$.  In addition, we observe a non-monotonic behavior in $R$ (see Fig.~\ref{fig.scatt}c and d), initially decreasing but then increasing with $f$. This behavior can be explained by the fact that quarantine has two opposite effects. On one hand, a quarantine strategy prevents the transmission of the disease, but on the other hand, it also induces fatigue in individuals, which will subsequently lead to an increase in the number of new cases.

\begin{figure}[H]
\vspace{0.0cm}
\begin{center}
\begin{overpic}[scale=0.35]{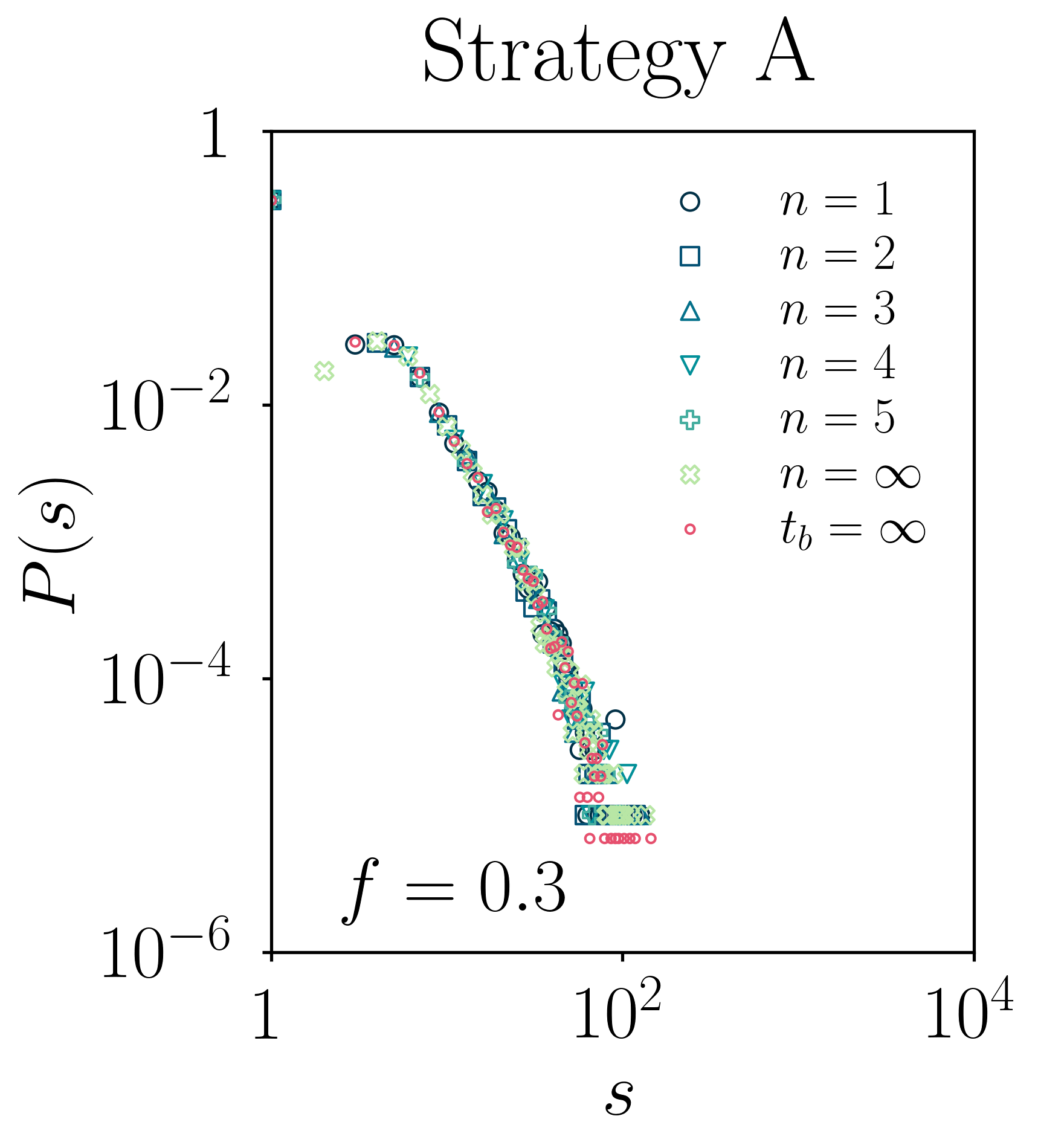}
  \put(30,35){(a)}
\end{overpic}
\vspace{0.0cm}
\begin{overpic}[scale=0.35]{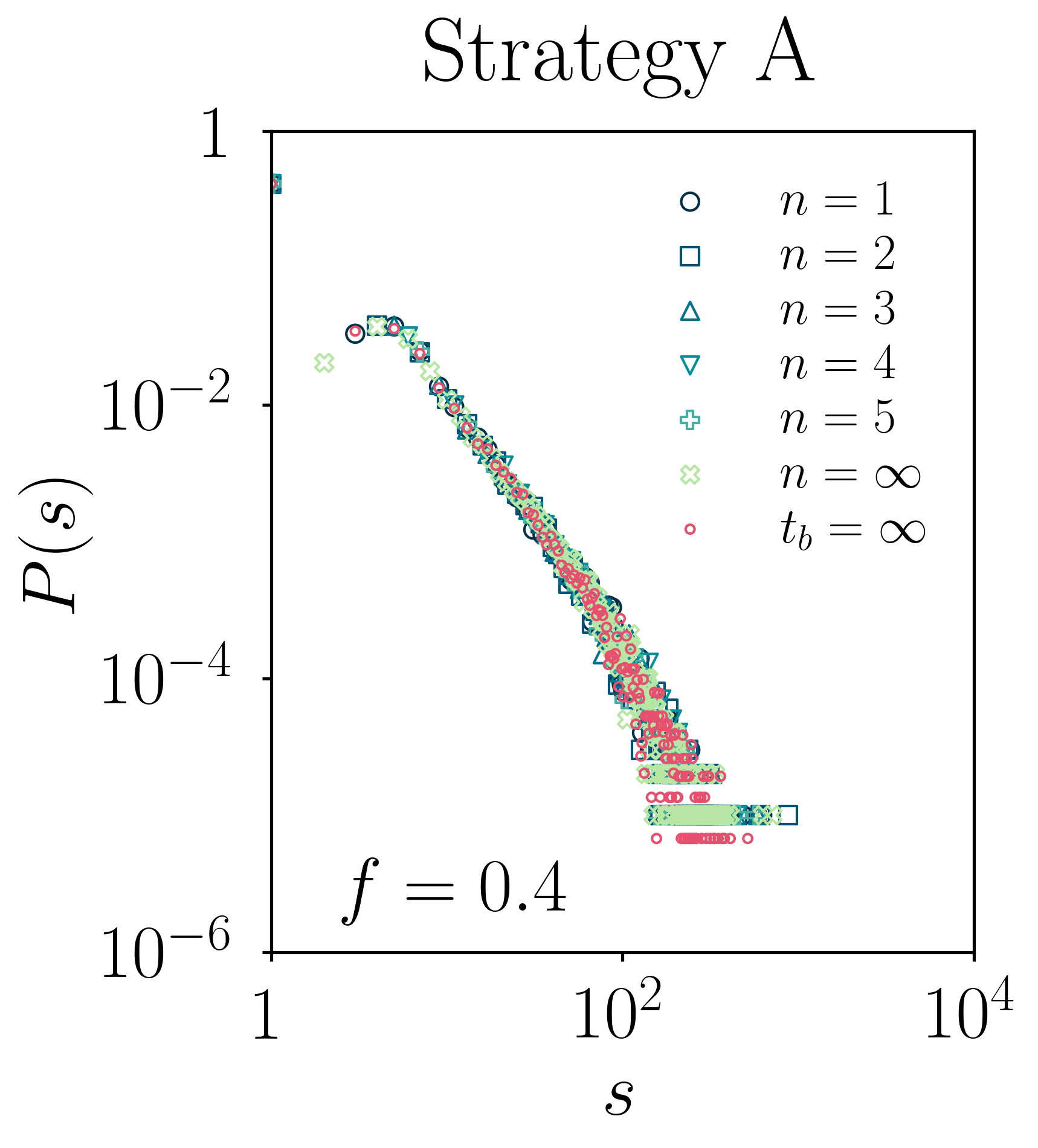}
  \put(30,35){(b)}
\end{overpic}
\vspace{0.0cm}
\begin{overpic}[scale=0.35]{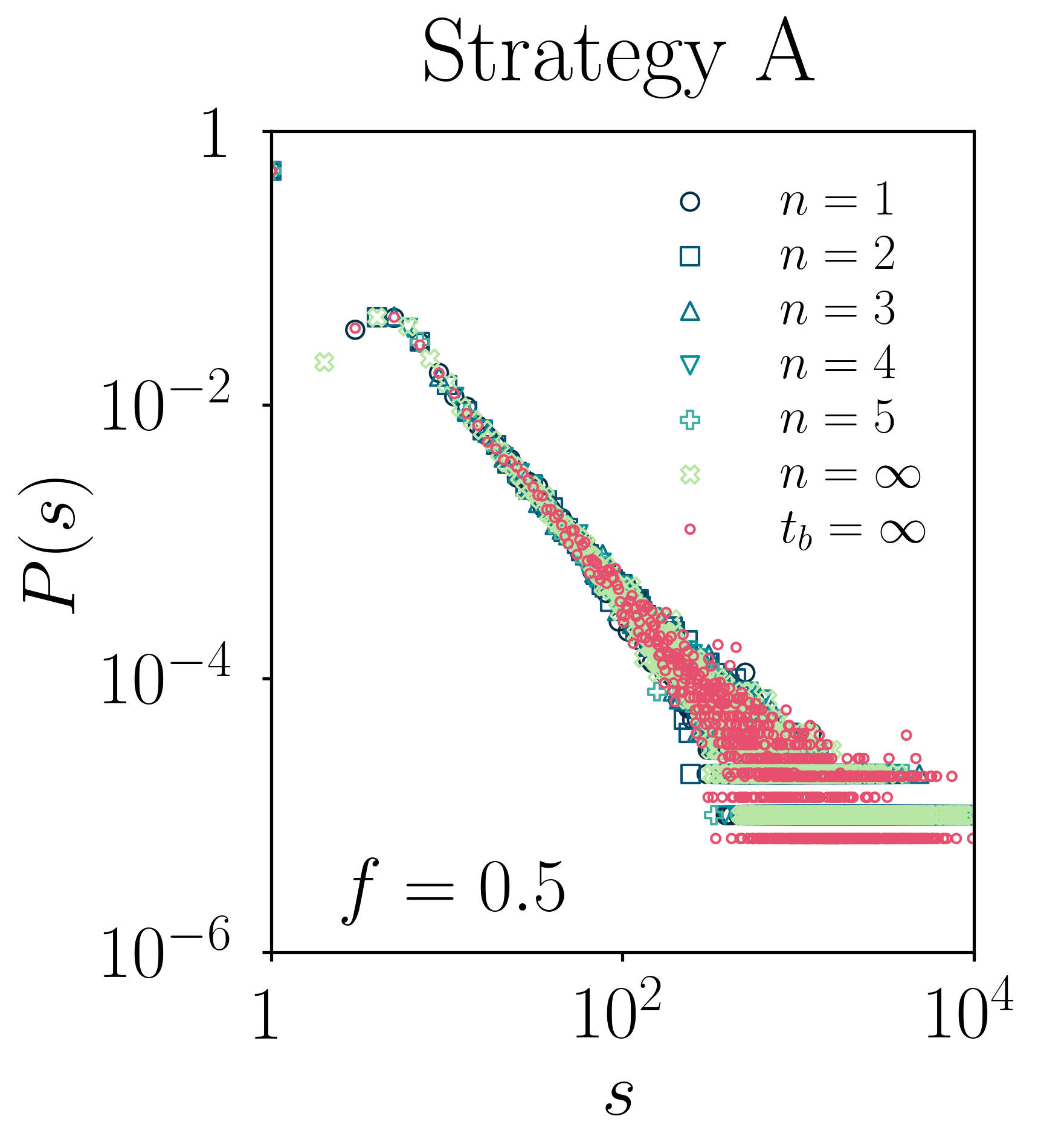}
  \put(30,35){(c)}
\end{overpic}
\begin{overpic}[scale=0.35]{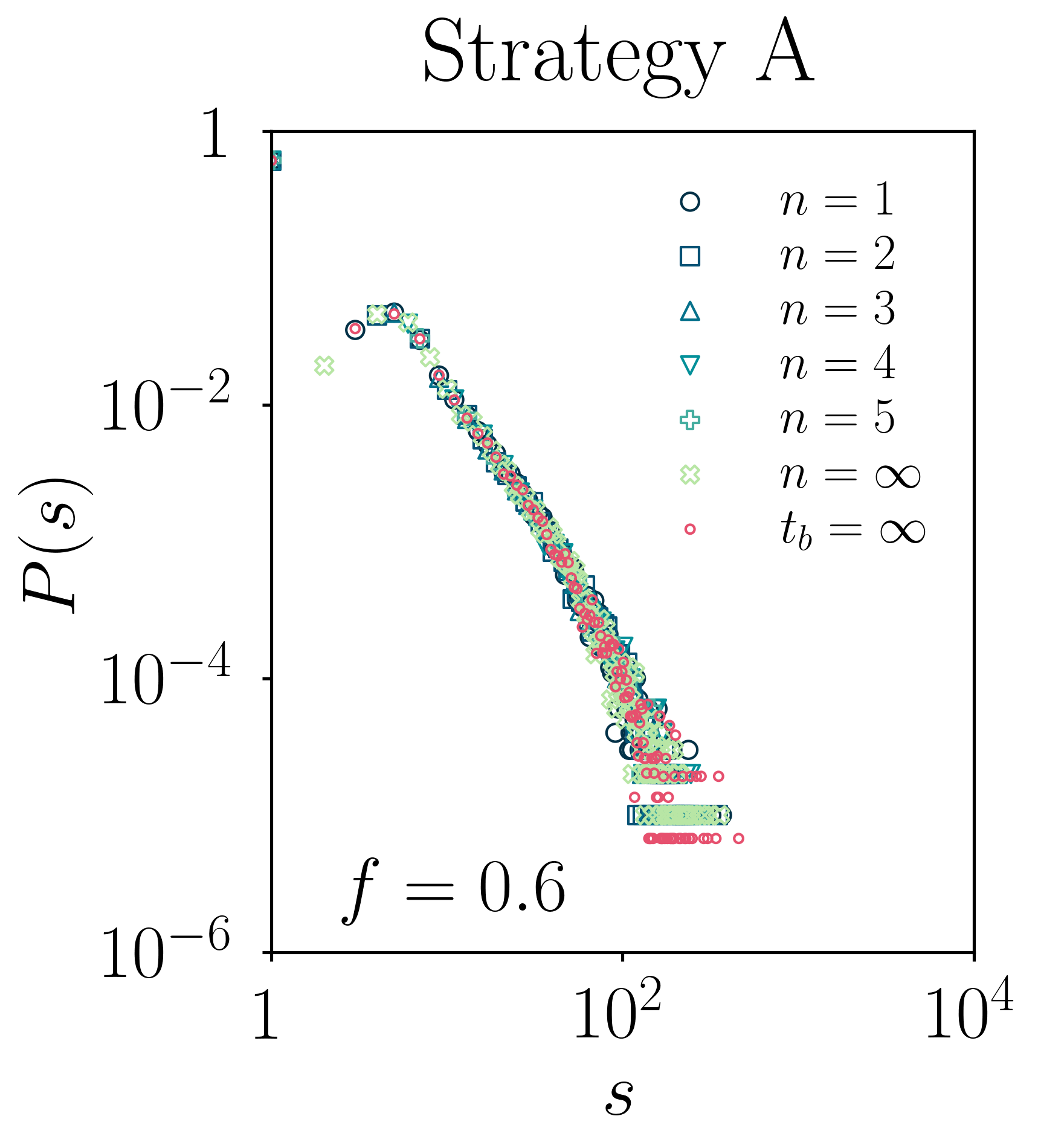}
  \put(30,35){(d)}
\end{overpic}
\begin{overpic}[scale=0.35]{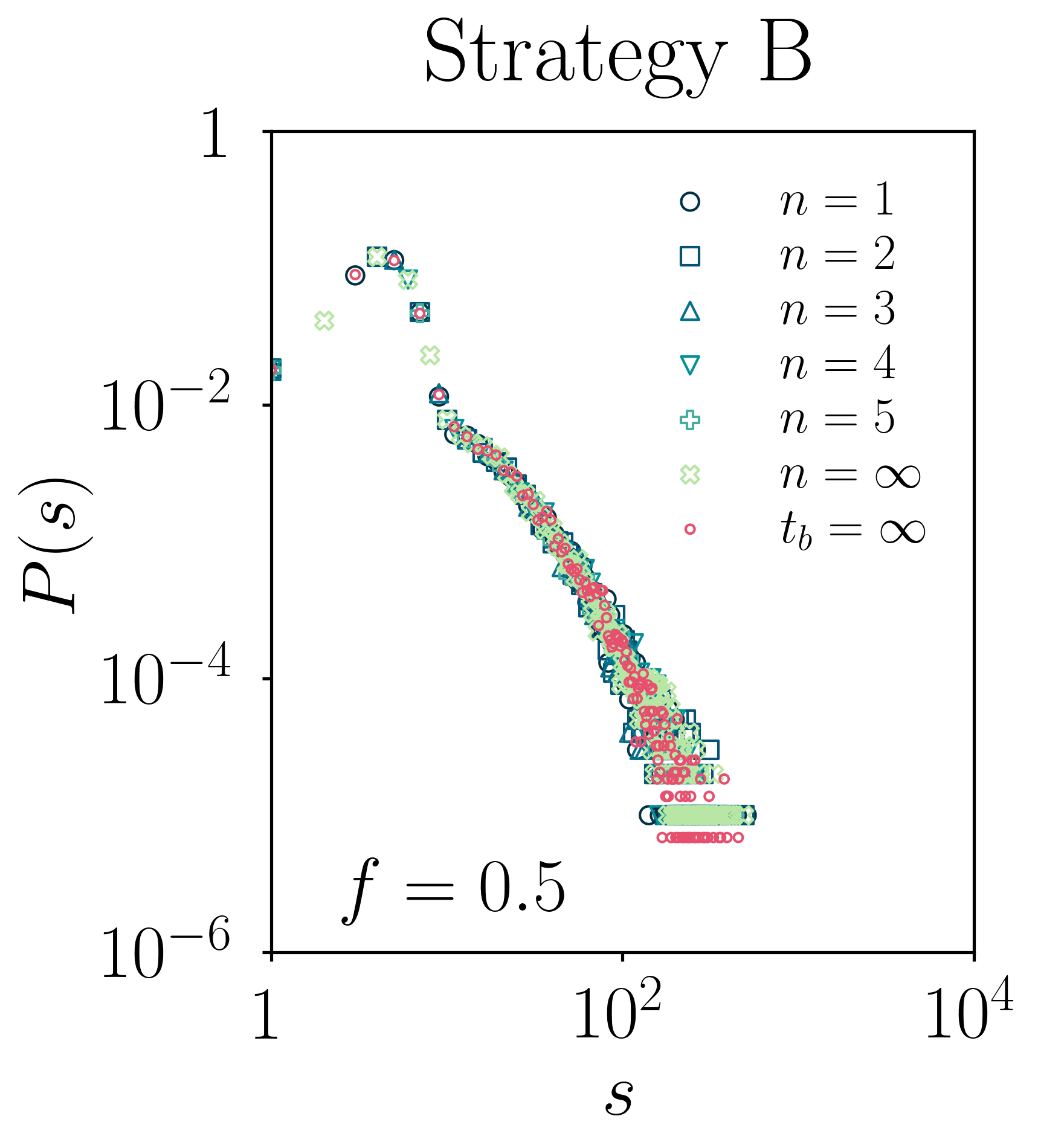}
  \put(30,35){(e)}
\end{overpic}
\vspace{0.0cm}
\begin{overpic}[scale=0.35]{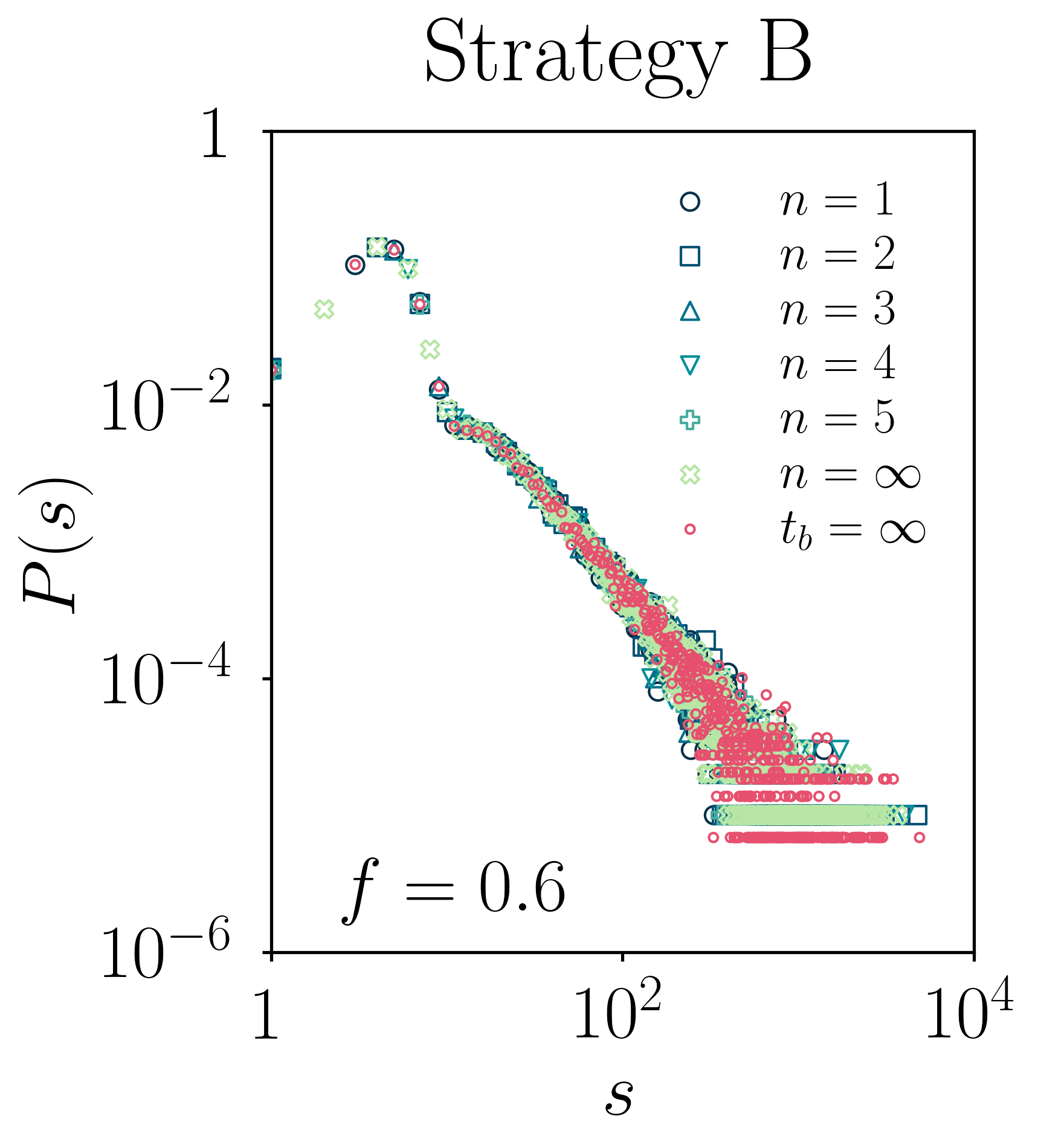}
  \put(30,35){(f)}
\end{overpic}
\vspace{0.0cm}
\begin{overpic}[scale=0.35]{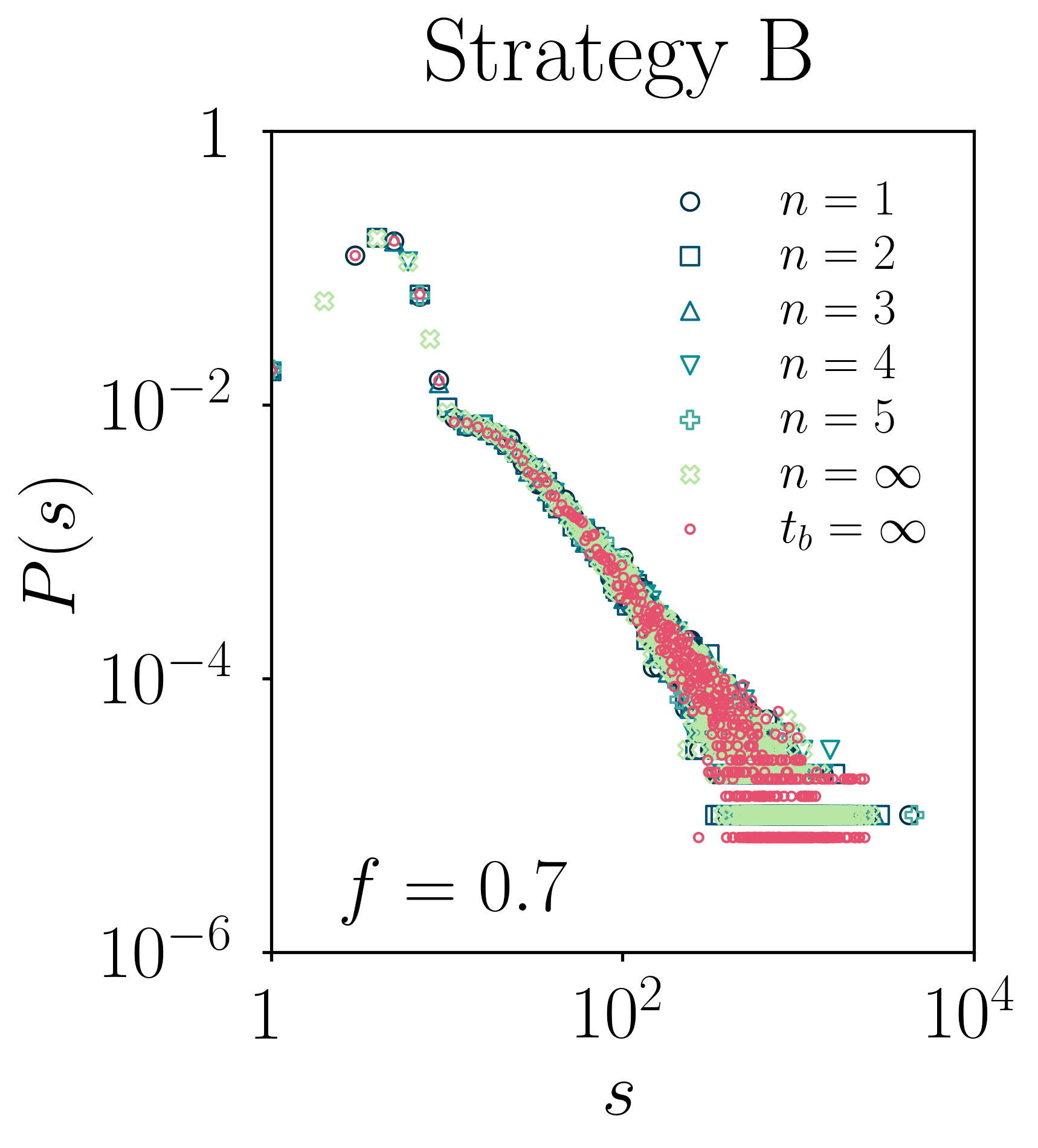}
  \put(30,35){(g)}
\end{overpic}
\begin{overpic}[scale=0.35]{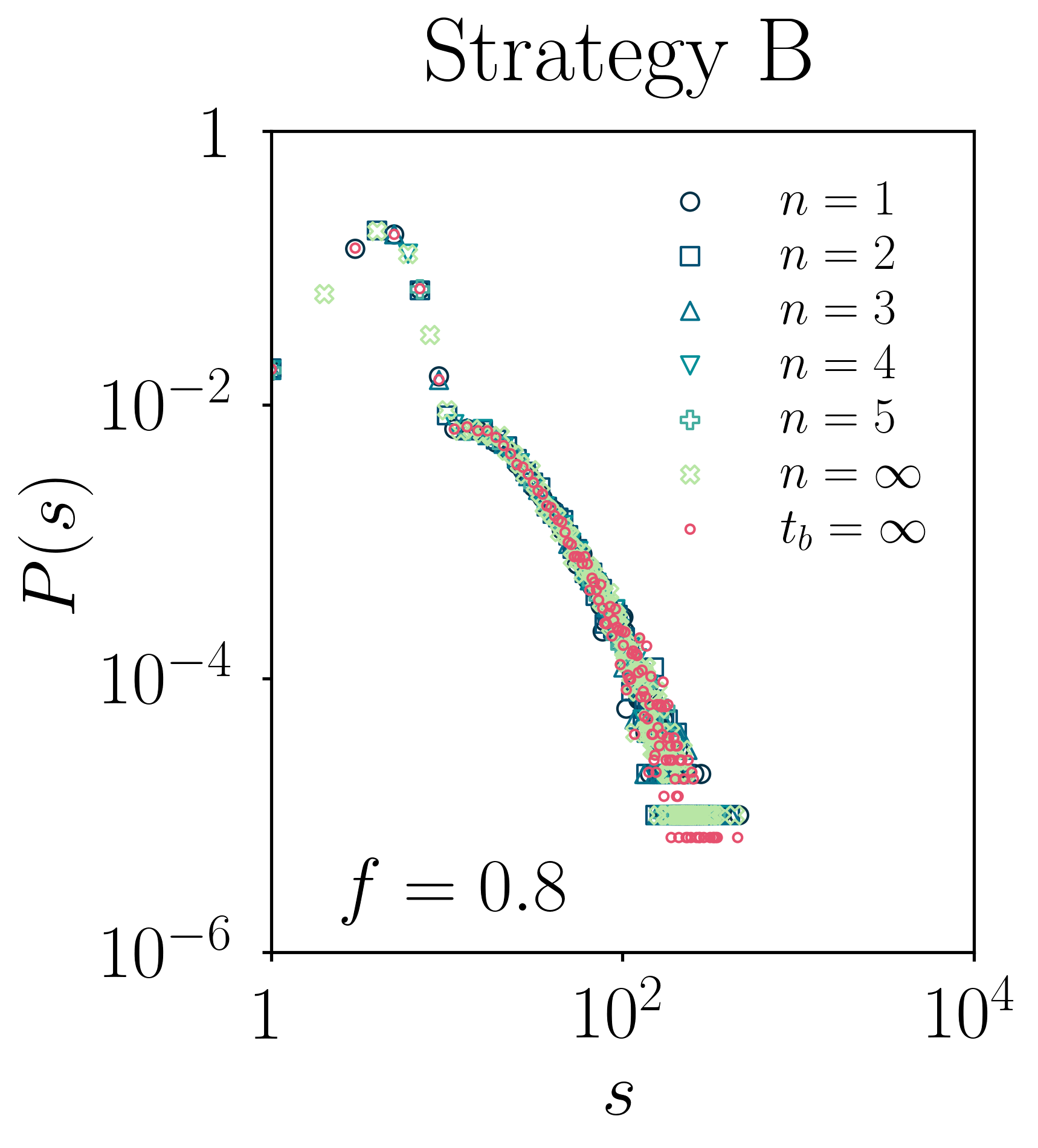}
  \put(30,35){(h)}
\end{overpic}
\vspace{0cm}
\vspace{-0.50cm}
\end{center}
\caption{Distribution $P(s)$ for Strategy A (panels a-d) and Strategy B (panels e-h) in log-log scale, with $\beta=0.20$, $N_I=10^6$, and various values of $f$ and $n$. Additionally, we also include the case $t_b=\infty$, where isolated individuals never rejoin the network. These results were obtained from $10^5$ stochastic realizations (symbols).}\label{fig.Ps}
\end{figure}

In addition to studying large-scale epidemics, it is also important to investigate small outbreaks because they provide valuable information on the critical behavior of the system around the transition point.  To this end, we compute the size distribution of small outbreaks, denoted as $P(s)$, for several values of $f$ and $n$; and display our results in Figs.~\ref{fig.Ps}a-h. First, from these figures, we can see that $P(s)$ decreases rapidly with $s$ when the detection probability $f$ is significantly distant from the transition point $f_c$. However, as $f$ approaches the vicinity of $f_c$, $P(s)$ behaves as a power-law function, suggesting that the system displays a second-order phase transition property. On the other hand, from these figures we can see that $P(s)$ does not show any dependence on the parameter $n$. This observation can be explained by considering that in a scenario where the number of infected people is microscopic (i.e., $s/N_I \ll 1$), the probability of an individual encountering multiple infected neighbors repeatedly is extremely low, and therefore, this individual will be isolated at most once and will not develop fatigue. Consequently, these results indicate that, for small outbreaks, our SIRQ model with fatigue is equivalent to our model explored in Ref.~\cite{valdez2023epidemic} where susceptible individuals do not return to the network and therefore do not experience fatigue.

In summary, our findings suggest that quarantine fatigue significantly influences the spread of diseases only when a substantial proportion of the population is infected. In addition, we obtain that for small outbreaks, the SIRQ model with fatigue is equivalent to the SIRQ model with $t_b=\infty$ presented in Ref.~\cite{valdez2023epidemic}. Therefore, the conclusions drawn in Ref.~\cite{valdez2023epidemic} for small outbreaks remain valid for our model with fatigue.

In the following section, we will show that the point around which the fraction of recovered individuals "explodes," corresponds to the basic reproduction number ($R_0$) equaling 1.

\section{Calculation of the basic reproduction number}\label{Sec.R0Calc}
One of the most commonly used measures in epidemiology is the basic reproduction number, denoted as $R_0$, which is defined as the average total number of individuals infected by a single patient at the beginning of the disease spread~\cite{anderson1991infectious,meyers2007contact}. In general, for values of $R_0$ less than 1 ($R_0 < 1$), the population remains free from epidemics. Conversely, when $R_0$ exceeds 1 ($R_0 > 1$), the likelihood of an epidemic becomes non-negligible. Finally, at the critical threshold of $R_0 = 1$, a second-order phase transition occurs, and above this threshold, the fraction of recovered individuals by the end of the epidemic tends to increase continuously as $R_0$ increases~\cite{anderson1991infectious}. It is worth noting, however, as previously indicated in the Introduction,  that certain models exhibit an explosive growth in the proportion of recovered individuals around $R_0 = 1$.

There are various methods to derive $R_0$. For instance, in mean-field compartmental models, the next-generation matrix method is employed~\cite{van2002reproduction,lindquist2011effective}. On the other hand, Miller~\cite{miller2009spread} proposed another approach to derive the basic reproduction number in complex networks using the concept of rank introduced by Ludwing~\cite{ludwig1975final}.

In the following, we will proceed to estimate the value of $R_0$ for both strategies using a method similar to that proposed by Miller~\cite{miller2009spread}.

\subsection{Estimation of $R_0$ for Strategy A}

In Sec.~\ref{sec.ResMicro}, we found that the transition point ($f_c$) is not affected by the parameters $n$ and $t_b$. This finding suggests that the basic reproduction number $R_0$ may also be independent of these parameters because typically, the transition point corresponds to the case where $R_0=1$. Consequently, we will use the formula of $R_0$ established in Ref.~\cite{valdez2023epidemic} (which corresponds to the SIRQ model with $t_b=\infty$) as an estimate for $R_0$ in our model with fatigue. The expression for $R_0$ (for $t_r=1$), as presented in Ref.~\cite{valdez2023epidemic}, is given by:
\begin{eqnarray}\label{eq.StA.R0}
R_0&=&\frac{\epsilon_1+\epsilon_2}{\mathcal{C}},
\end{eqnarray}
where $\mathcal{C}$, $\epsilon_1$ and $\epsilon_2$ are given by:
\begin{eqnarray}
  \mathcal{C}&=&\beta(k_C-1),\label{eq.CStratA}\\
  \epsilon_1&=&(k_C-1)(1-\beta)\left[(\beta(1-f)+(1-\beta))^{k_C-2}-(\beta(1-\beta)(1-f)+(1-\beta))^{k_C-2}\right],\label{eq.e1StratA}\\
  \epsilon_2&=&(1-f)(1-\beta f)^{k_C-2}(k_I-1)(k_C-1)^2\beta^2.\label{eq.e2StratA}
\end{eqnarray}
Note that Eqs.~(\ref{eq.CStratA})-(\ref{eq.e2StratA}) only apply specifically to cases where all cliques have the same number of individuals ($k_C$), and each person belongs to an identical number of cliques ($k_I$). In Eq.~(\ref{eq.StA.R0}), the denominator accounts for the number of infections directly caused by an index case within a clique. On the other hand, the numerator $\epsilon_1+\epsilon_2$, represents the sum of individuals in the second generation of infections. Here, $\epsilon_1$ refers to those individuals at a chemical distance~\cite{cohen2010complex} of $\ell = 1$ from the index case, while $\epsilon_2$ refers to those individuals at a chemical distance of $\ell = 2$ from the index case.

In Sec.~\ref{sec.R0Results}, we will see that Eq.~(\ref{eq.StA.R0}) is valid for our model with fatigue on random regular networks with cliques. Furthermore, using the condition $R_0=1$, we will investigate the relationship between the critical detection probability $f_c$ and the clique size $k_C$ (see Appendix~\ref{App.fckc}).

\subsection{Estimation of $R_0$ for Strategy B}
Following a similar line of reasoning as in Strategy A, we find that for Strategy B, the basic reproduction number, for the case of random regular networks with cliques, is given by:
\begin{eqnarray}\label{eq.StB.R0}
R_0&=&\frac{\epsilon_1+\epsilon_2}{\mathcal{C}},
\end{eqnarray}
where $\mathcal{C}$, $\epsilon_1$ and $\epsilon_2$ are given by:
\begin{eqnarray}
  \mathcal{C}&=&\beta(k_C-1),\label{eq.CStratB}\\
  \epsilon_1&=&(k_C-1)(1-\beta)\left[(\beta(1-f)+(1-\beta))^{k_C-2}-(\beta(1-\beta)(1-f)+(1-\beta))^{k_C-2}\right],\label{eq.e1StratB}\\
  \epsilon_2&=&(1-f)(k_I-1)(k_C-1)^2\beta^2.\label{eq.e2StratB}
\end{eqnarray}
Note that Eq.~(\ref{eq.CStratA}) and Eq.~(\ref{eq.CStratB}) are identical, as well as Eq.~(\ref{eq.e1StratA}) and Eq.~(\ref{eq.e1StratB}). On the other hand, the only distinction between Eq.~(\ref{eq.e2StratA}) and Eq.~(\ref{eq.e2StratB}) lies in the factor $(1-\beta f)^{k_C-2}$ which represents the probability that an individual has not been isolated by any of their infected neighbors.

In the following section, we will compare Eq.~(\ref{eq.StB.R0}) with our numerical simulations. Furthermore, in Appendix~\ref{App.fckc}, we investigate the relationship between the critical detection probability $f_c$ and the clique size $k_C$.

\subsection{Comparison with Numerical Results}\label{sec.R0Results}

\begin{figure}[H]
\vspace{0.0cm}
\begin{center}
\begin{overpic}[scale=0.29]{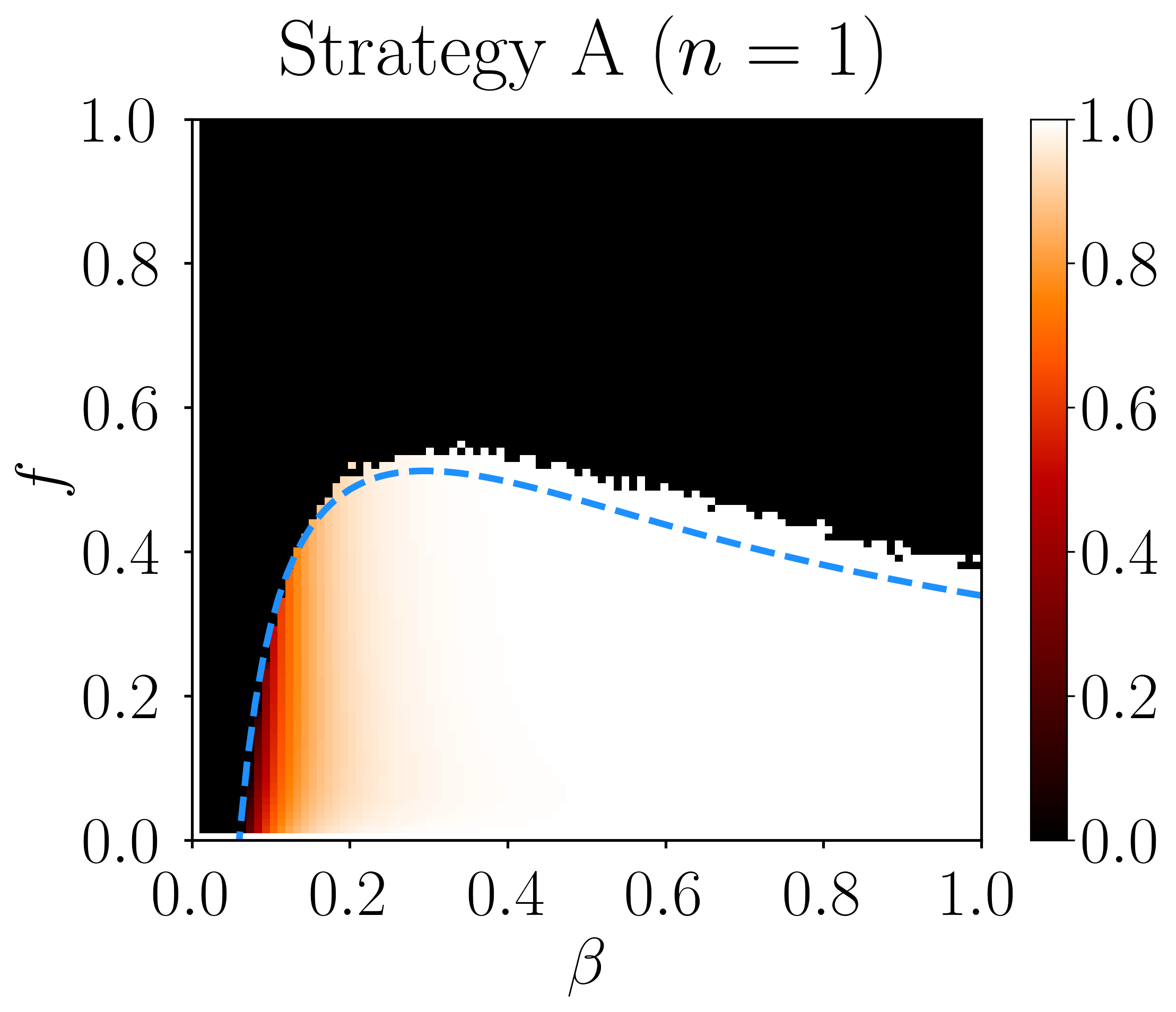}
  \put(70,25){(a)}
\end{overpic}
\vspace{0.0cm}
\begin{overpic}[scale=0.29]{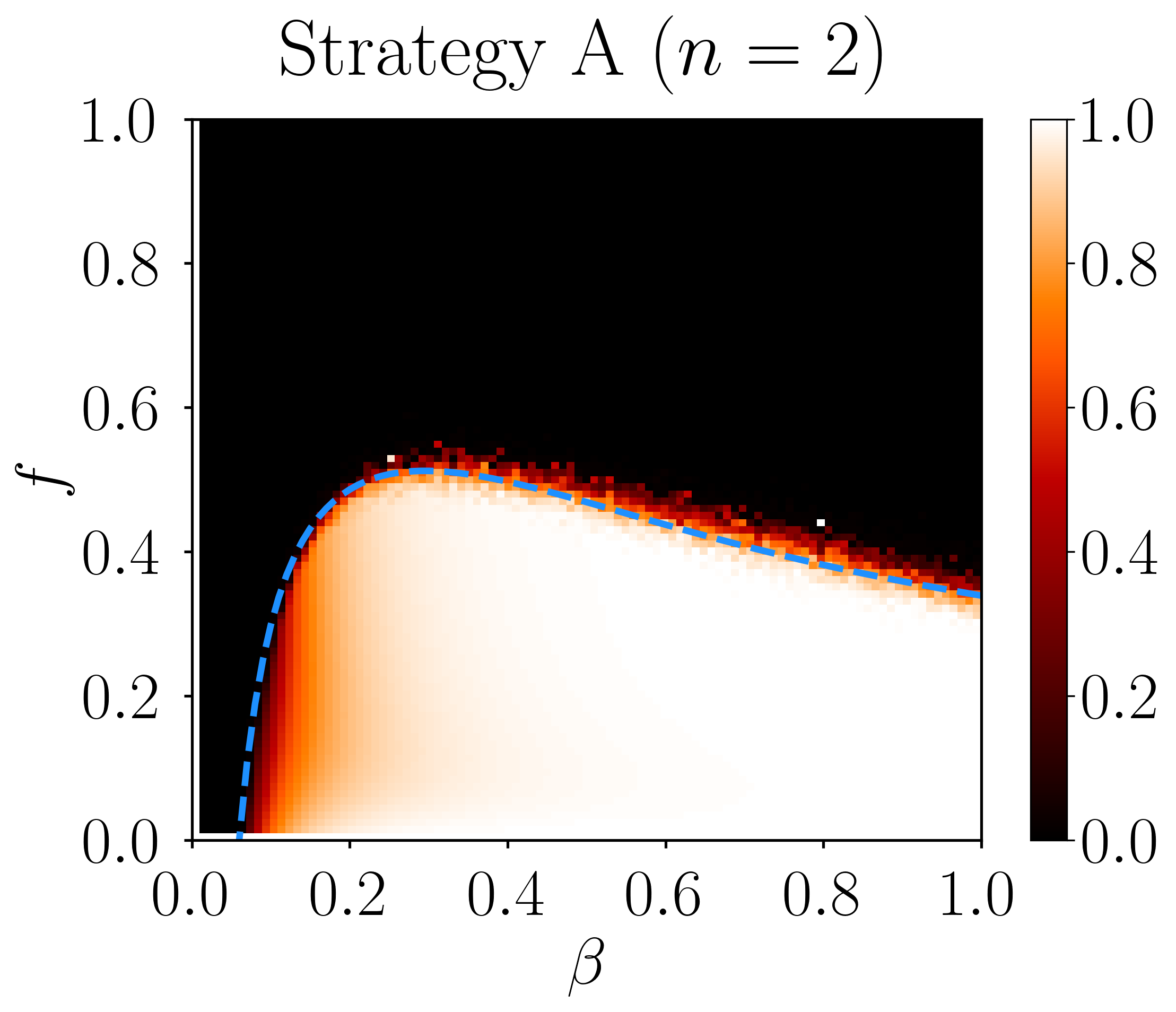}
  \put(70,25){(b)}
\end{overpic}
\vspace{0.0cm}
\begin{overpic}[scale=0.29]{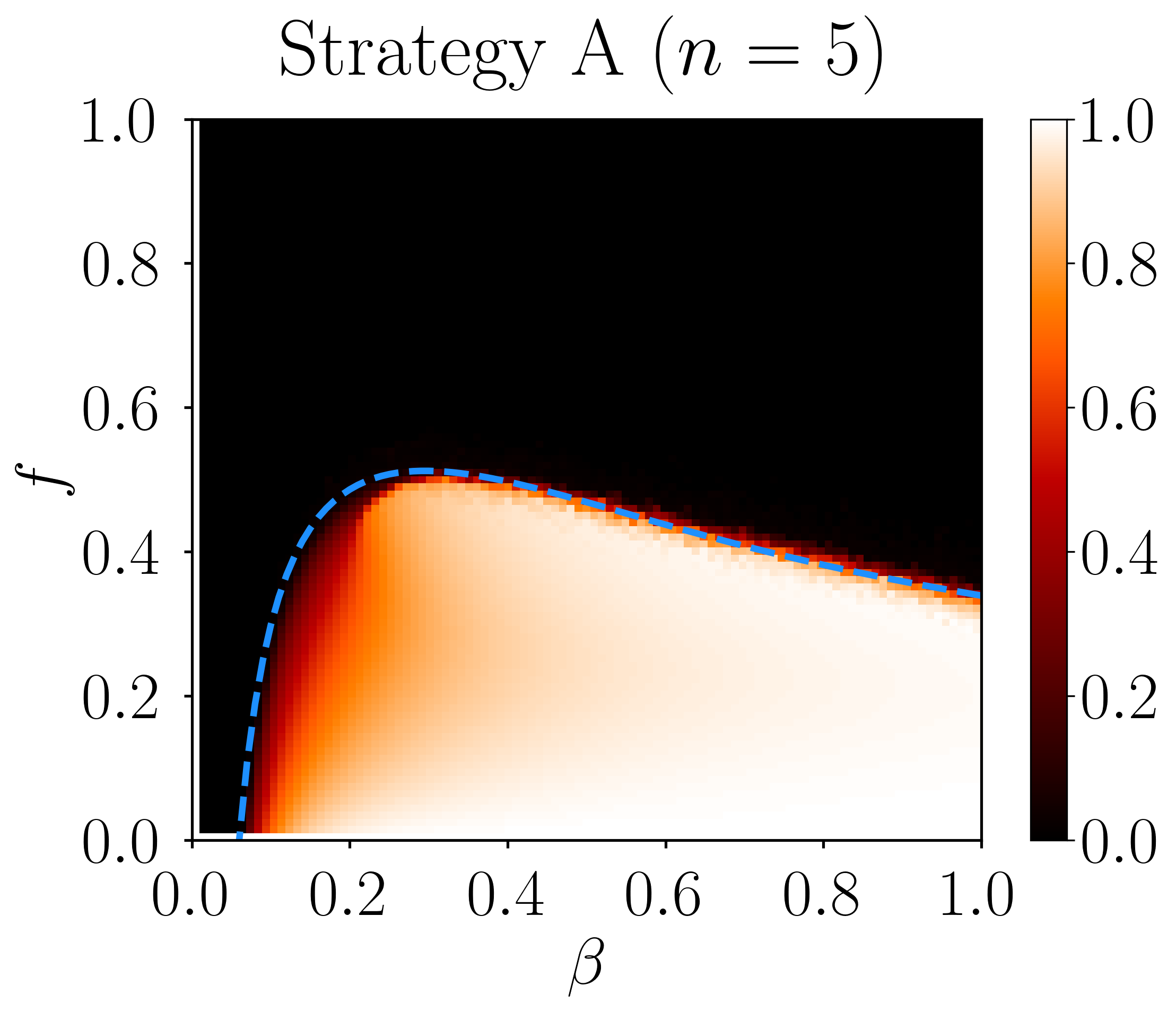}
  \put(70,25){(c)}
\end{overpic}
\begin{overpic}[scale=0.29]{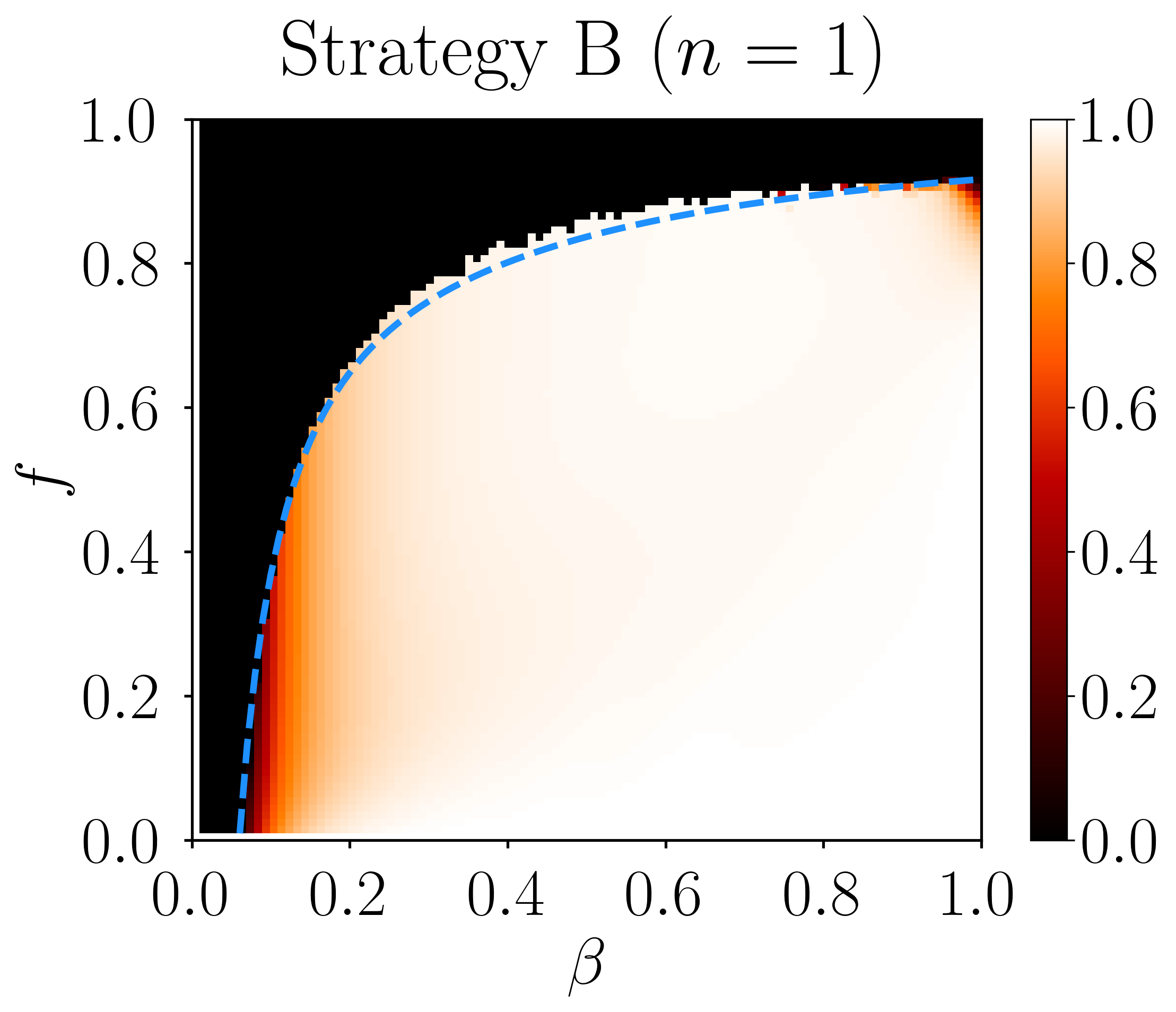}
  \put(70,25){(d)}
\end{overpic}
\vspace{0.0cm}
\begin{overpic}[scale=0.29]{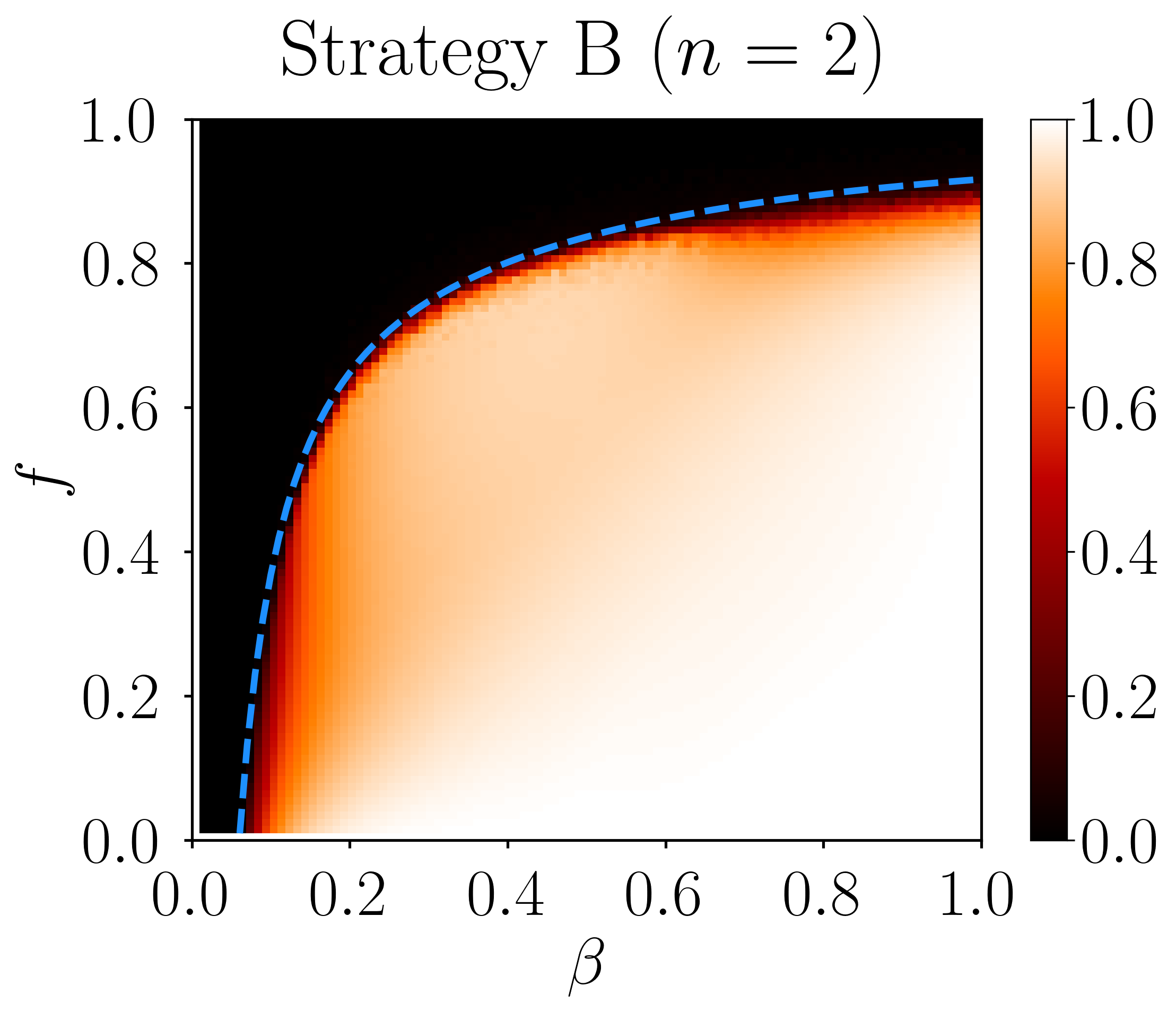}
  \put(70,25){(e)}
\end{overpic}
\vspace{0.0cm}
\begin{overpic}[scale=0.29]{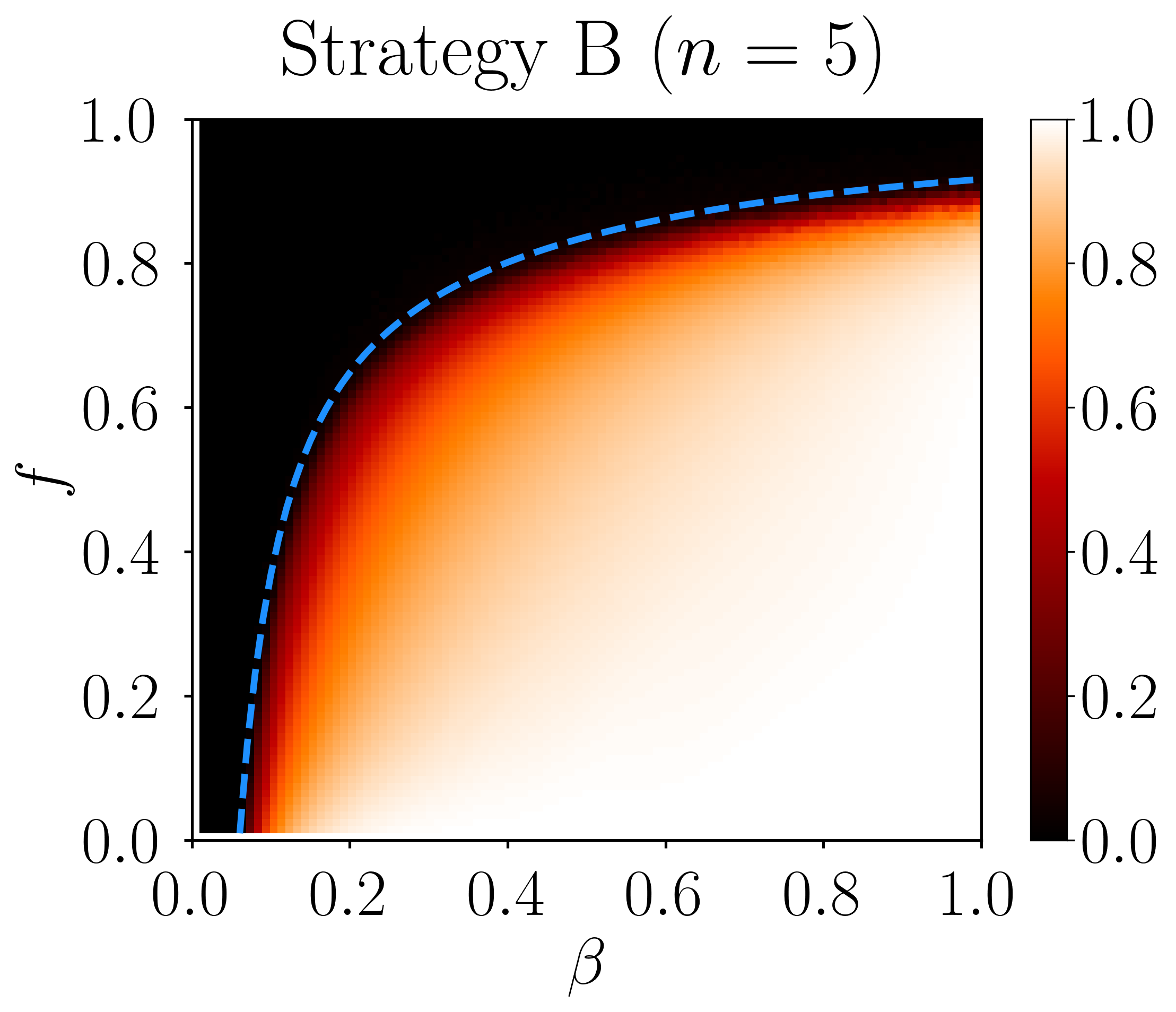}
  \put(70,25){(f)}
\end{overpic}
\vspace{-0.60cm}
\end{center}
\caption{Heat-maps in the $\beta-f$ plane showing the fraction of recovered individuals at the final stage for Strategy A (panels a-c) and Strategy B (panels d-f), with $t_r=1$, $t_b=1$ and considering different values of $n$. Results were obtained from $10^3$ stochastic realizations on RR networks with $k_C=7$, $k_I=3$. For $n>1$, simulations were conducted on networks with $N=10^5$. However, for $n=1$, a larger network size of $N=2\times 10^6$ was used to mitigate finite-size effects and reduce simulation noise. Blue dashed lines represent the critical curves obtained from Eqs.~(\ref{eq.StA.R0}) and (\ref{eq.StB.R0}). }\label{fig.betfplane}
\end{figure}

To confirm the validity of Eqs.~(\ref{eq.StA.R0}) and (\ref{eq.StB.R0}), in Figs.~\ref{fig.betfplane}a-f we present the fraction of recovered population $R$ at the final stage in the $\beta-f$ plane obtained from our stochastic simulations for random regular networks with cliques (with $k_C=7$ and $k_I=3$) and various values of $n$. In addition, each figure includes a curve representing the condition $R_0 = 1$ derived from Eqs.~(\ref{eq.StA.R0}) and (\ref{eq.StB.R0}), which we refer to as the critical curve.

On one hand, Figs.~\ref{fig.betfplane}a-f reveal a smaller epidemic region for Strategy A compared to Strategy B, as expected, since the former corresponds to an early intervention where individuals are detected before they transmit the disease, while in the latter the disease can be transmitted and "escape" before the clique is quarantined. On the other hand, these figures demonstrate that each critical curve closely delineates the boundary between the epidemic phase and the disease-free phase. As a result, for both Strategy A and Strategy B, a phase transition occurs at $R_0=1$ and does not depend on the value of $n$ when the initial condition is microscopic. However, it is important to note that the nature of this transition---continuous or discontinuous--- does depend on the values of $\beta$, $t_b$, and $n$. 

In the following section, we will explore the influence of the fatigue threshold on the transition point for a scenario where a non-negligible number of individuals are infected at the beginning of an epidemic.

\section{Results for a macroscopic initial condition}\label{sec.ResMacro}

In the preceding sections, our focus was primarily on the case in which a single infected person initiates a disease outbreak. However, it is also important to consider scenarios with a substantial number of initially infected individuals. This could, for instance, represent cases where control measures are implemented after a substantial portion of the population is already infected~\cite{radicchi2020epidemic}. Additionally, it was observed that some systems exhibiting a discontinuous transition have a strong dependency on initial conditions~\cite{ferraz2023multistability,tuzon2018continuous,chen2019nontrivial}. In this section, we will explore how our model behaves when the initial proportion of infected individuals ($I_0$) is macroscopic.

\begin{figure}[H]
\begin{center}
\begin{overpic}[scale=0.29]{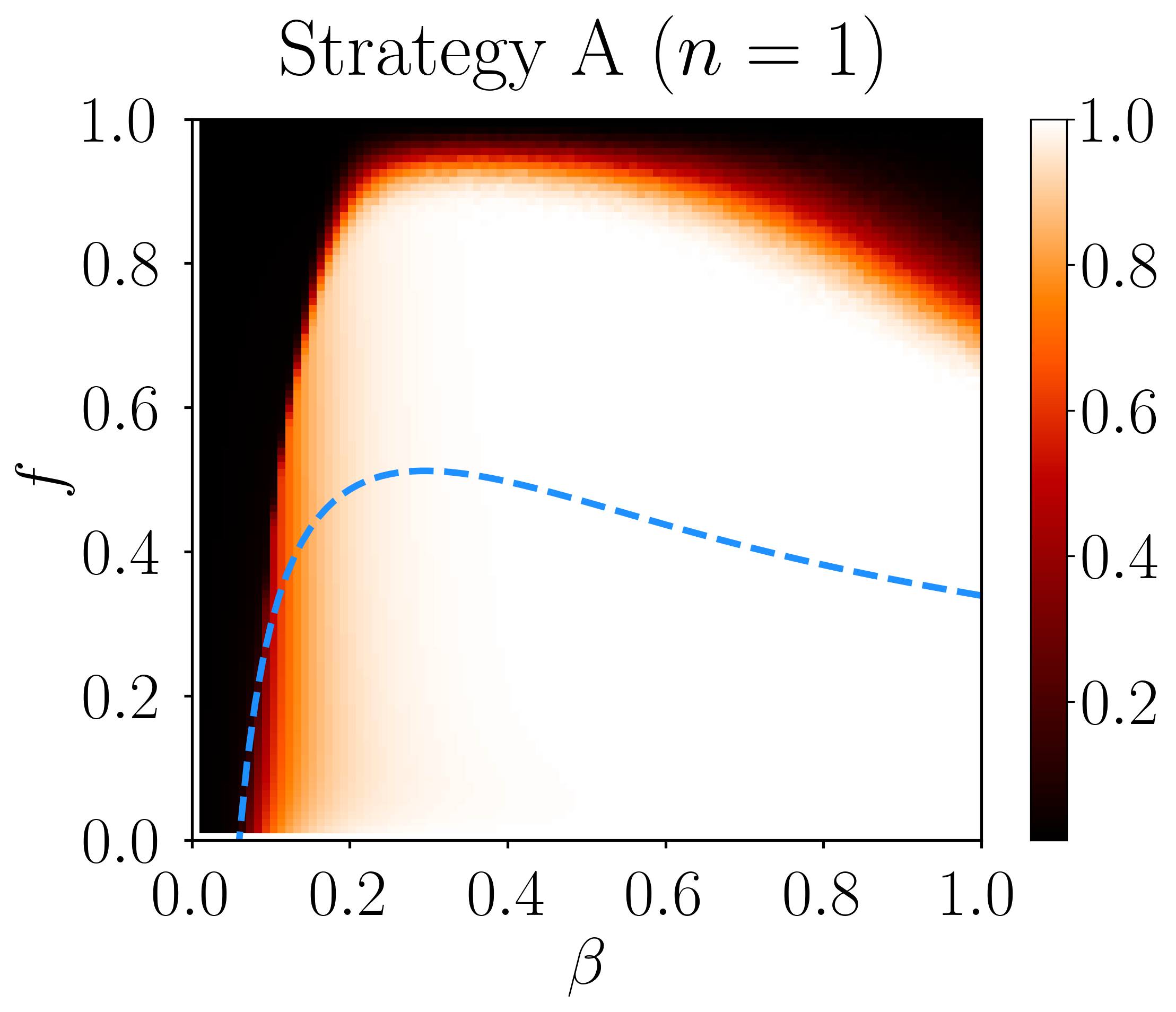}
  \put(70,25){(a)}
\end{overpic}
\begin{overpic}[scale=0.29]{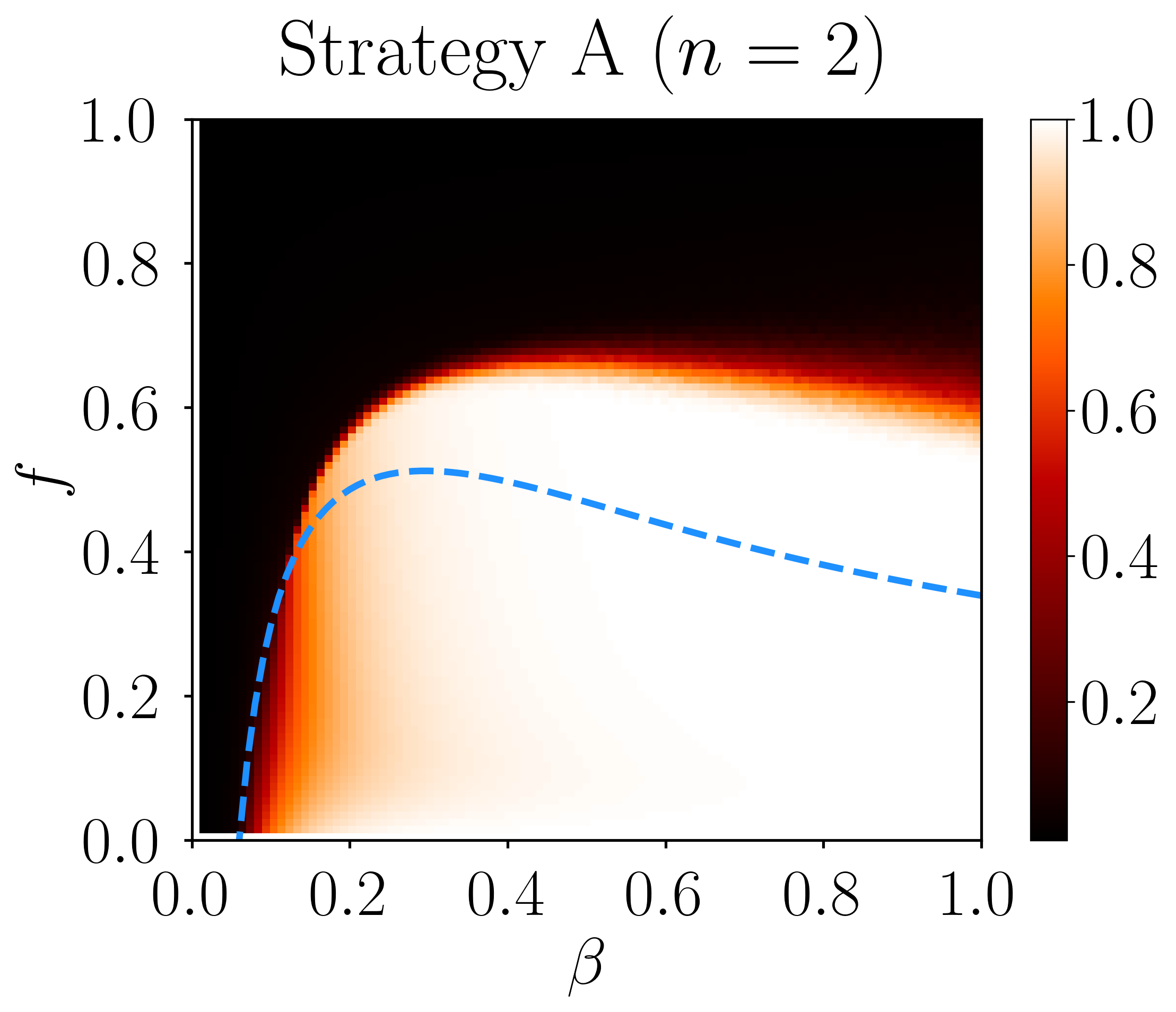}
  \put(70,25){(b)}
\end{overpic}
\begin{overpic}[scale=0.29]{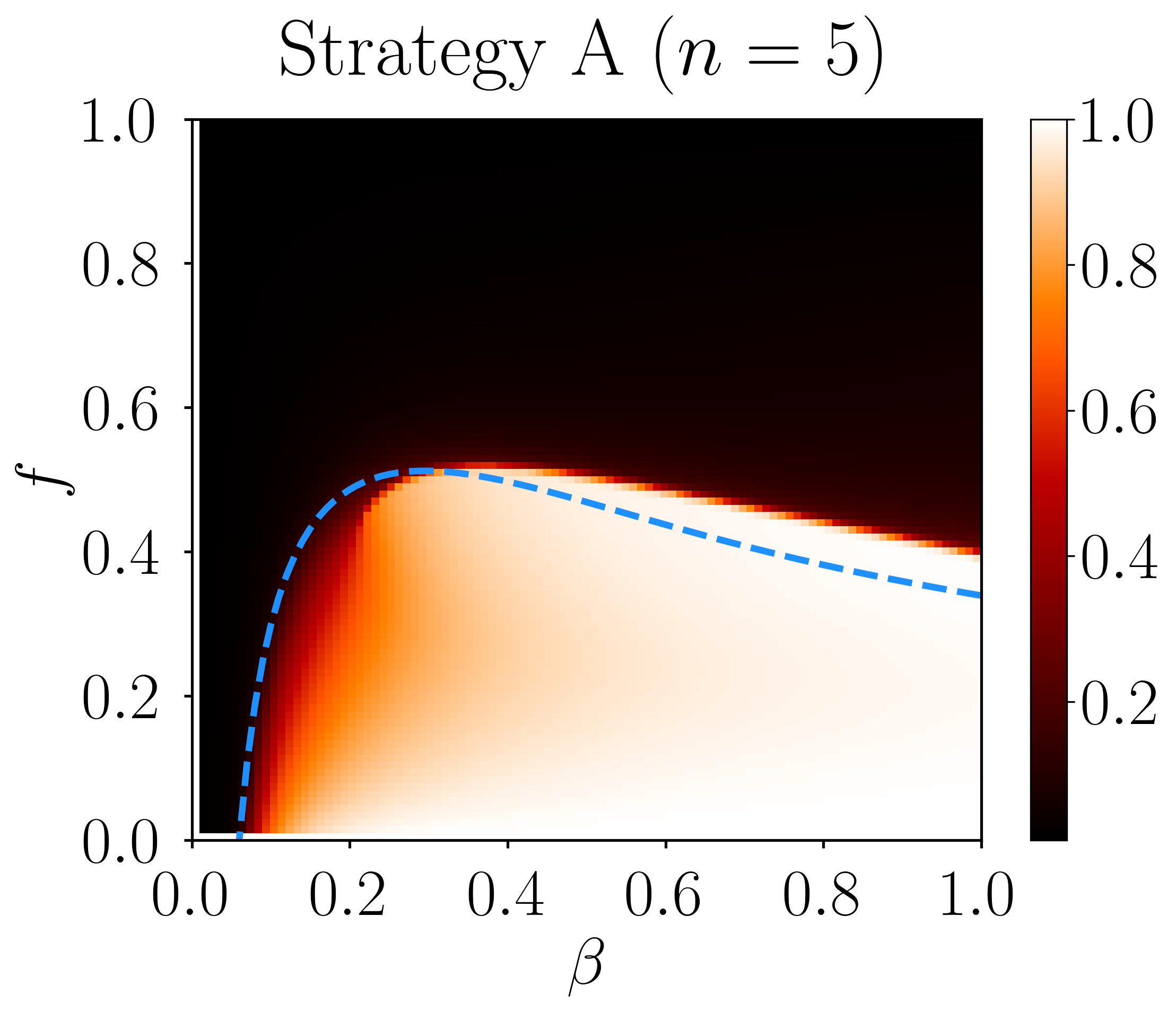}
  \put(70,25){(c)}
\end{overpic}
\begin{overpic}[scale=0.29]{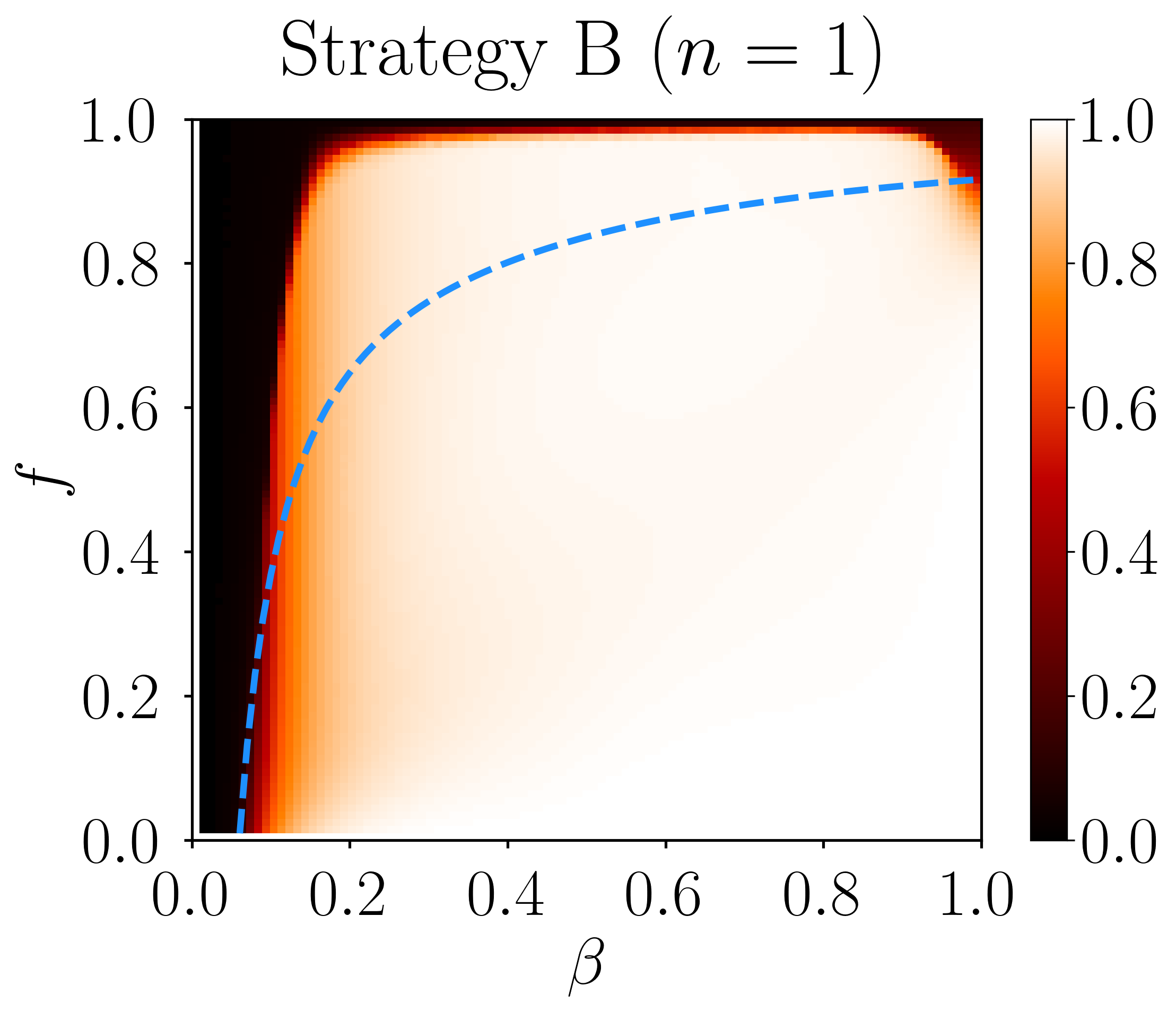}
  \put(70,25){(d)}
\end{overpic}
\begin{overpic}[scale=0.29]{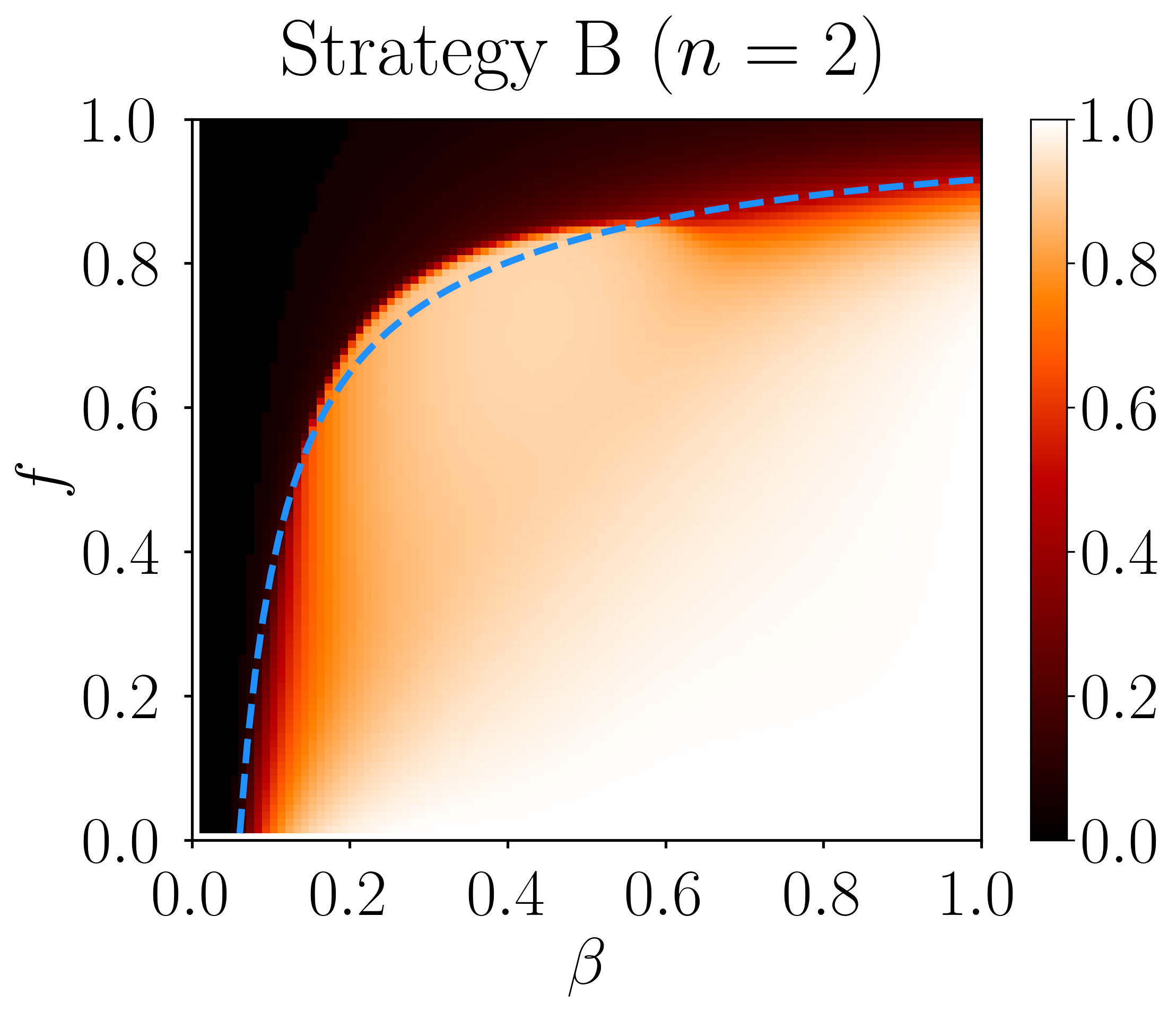}
  \put(70,25){(e)}
\end{overpic}
\begin{overpic}[scale=0.29]{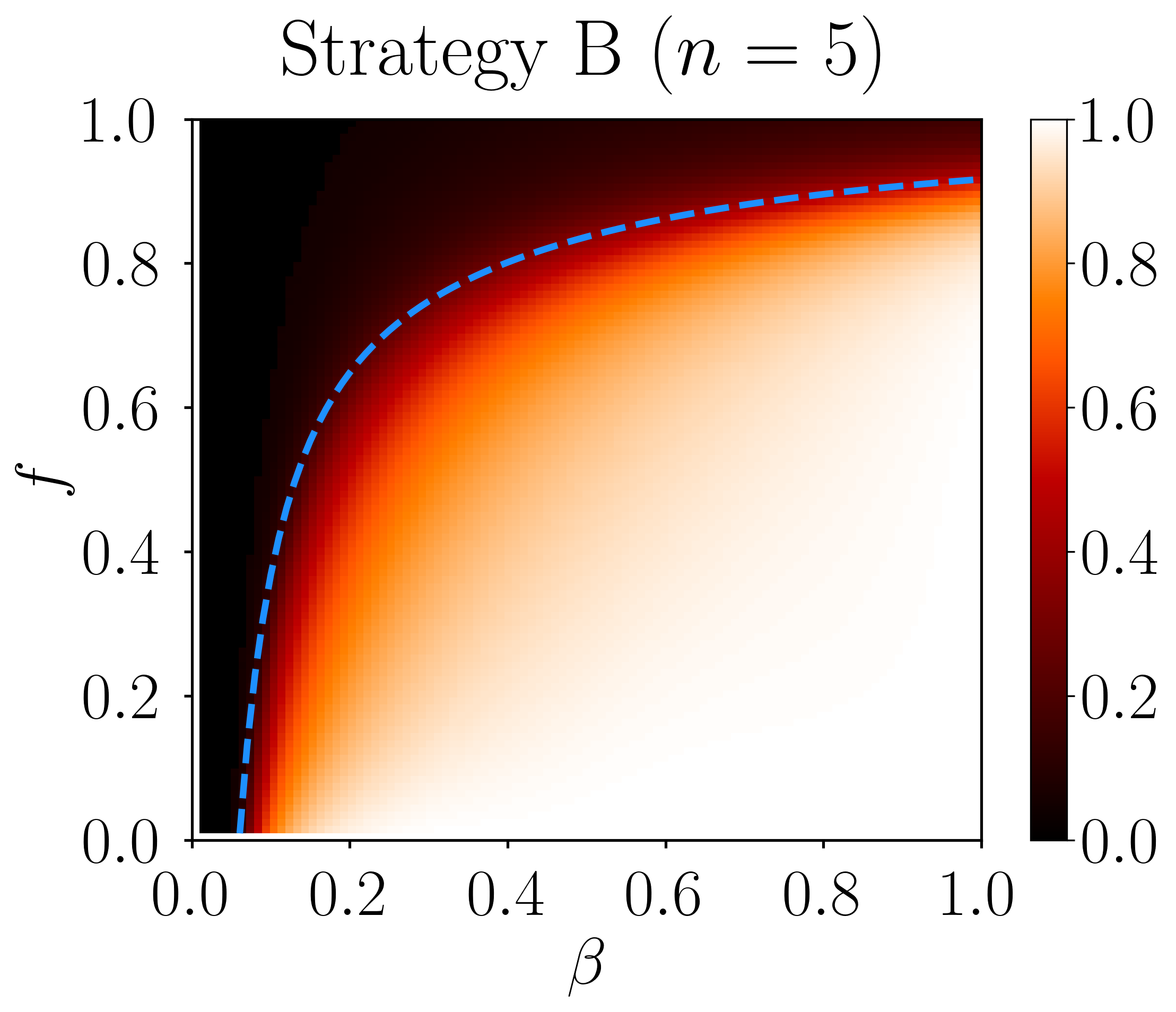}
  \put(70,25){(f)}
\end{overpic}
\vspace{-0.6cm}
\end{center}
\caption{Heat-maps in the $\beta-f$ plane showing the fraction of recovered individuals at the final stage for Strategy A (panels a-c) and Strategy B (panels d-f), with $I_0=1$\%, with $t_r=1$, $t_b=1$ and considering different values of $n$. Results were obtained from $10^3$ stochastic realizations on random regular networks with $k_C=7$, $k_I=3$, and $N_I=10^5$. For comparison, we also plot the critical curves (blue dashed lines) obtained from Eqs.~(\ref{eq.StA.R0}) and (\ref{eq.StB.R0}) which correspond to a microscopic initial condition. }\label{fig.betfplaneI0}
\end{figure}

Figures~\ref{fig.betfplaneI0}a-f show the fraction of recovered population $R$ at the final stage in the $\beta-f$ plane obtained from our stochastic simulations for Strategy A and Strategy B, and $I_0=1$\%. Note that for a macroscopic initial condition, the final stage always corresponds to an epidemic because $R\geq I_0$=1\%, or in other words, the epidemic-free phase does not exist. As a consequence, there is no critical threshold curve separating an epidemic phase from an epidemic-free one. However, from Figs.~\ref{fig.betfplaneI0} a-e,  we still observe a discernible threshold curve around which $R$ undergoes an abrupt transition, i.e., the disease goes from a controlled epidemic ($R\gtrsim 1$\%) to an explosive epidemic where approximately almost the entire population is infected. Notably, the position of this threshold curve does depend on the fatigue threshold, which contrasts with the case where the initial number of infected people is very small, shown in Sec.~\ref{sec.R0Results}.

To further explore the impact of initial conditions on our model, we will next study $R$ as a function of $I_0$. Our focus will be in particular on scenarios where the basic reproduction number $R_0$ is less than 1 because, for $R_0$ values greater than 1, an epidemic can emerge even from a single infected person at the beginning of the disease outbreak. In other words, we will choose the values of $\beta$ and $f$ such that $R_0<1$ (using Eqs.~(\ref{eq.StA.R0})-(\ref{eq.e2StratA}) for Strategy A and Eqs.~(\ref{eq.StB.R0})-(\ref{eq.e2StratB}) for Strategy B).

\begin{figure}[H]
\vspace{0.0cm}
\begin{center}
\begin{overpic}[scale=0.45]{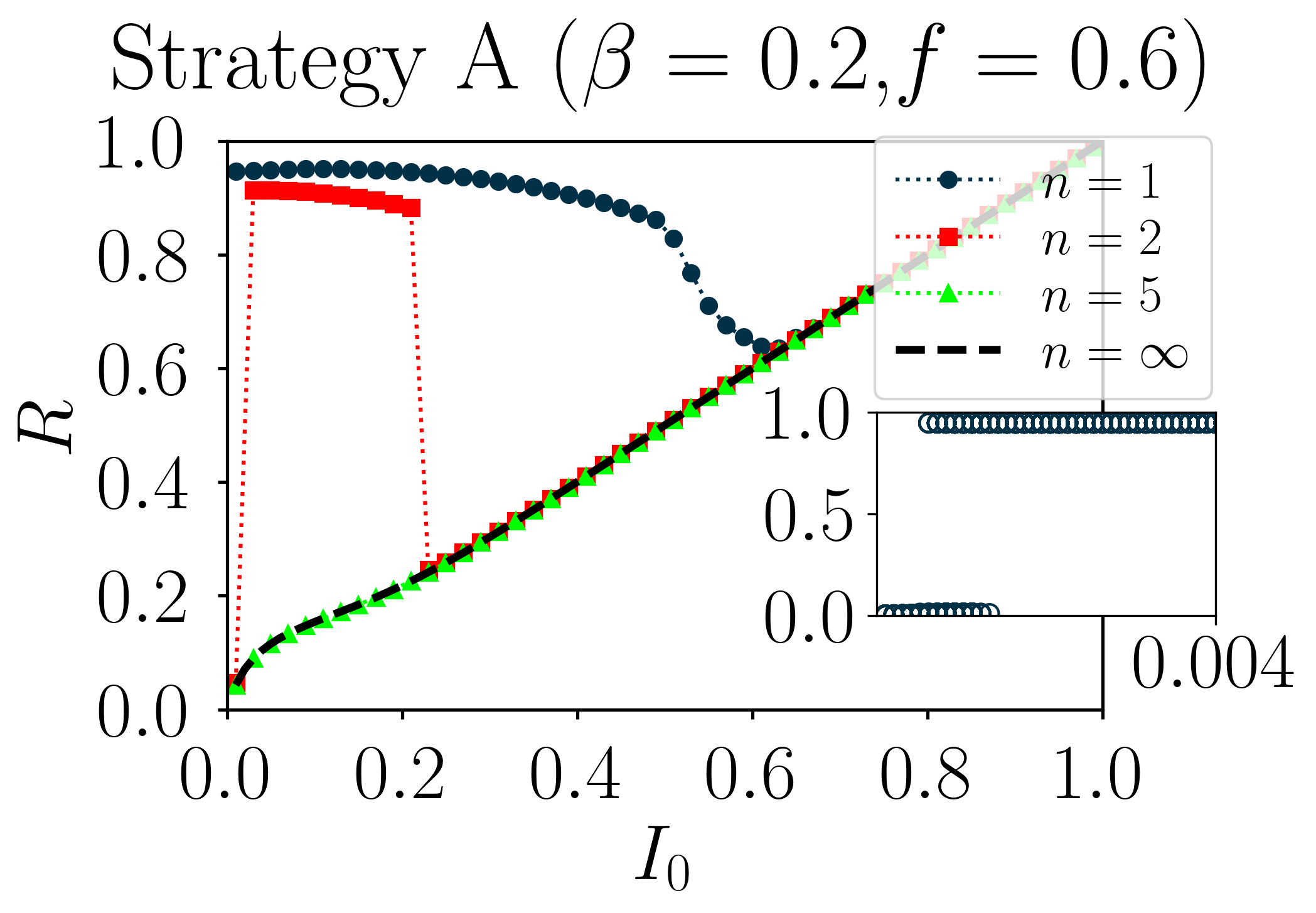}
  \put(30,20){(a)}
\end{overpic}
\vspace{0.0cm}
\begin{overpic}[scale=0.45]{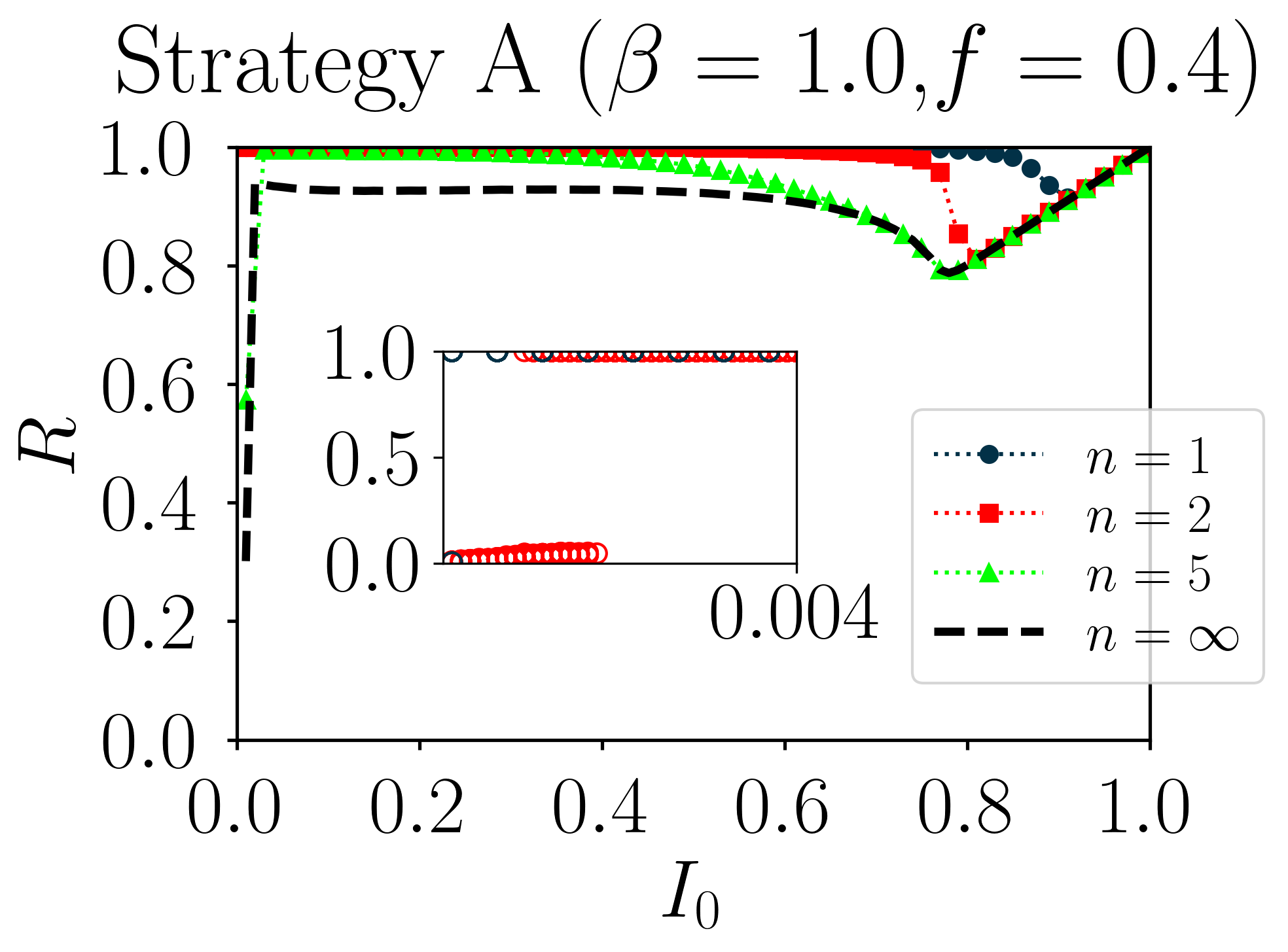}
  \put(30,20){(b)}
\end{overpic}
\vspace{0.0cm}
\begin{overpic}[scale=0.45]{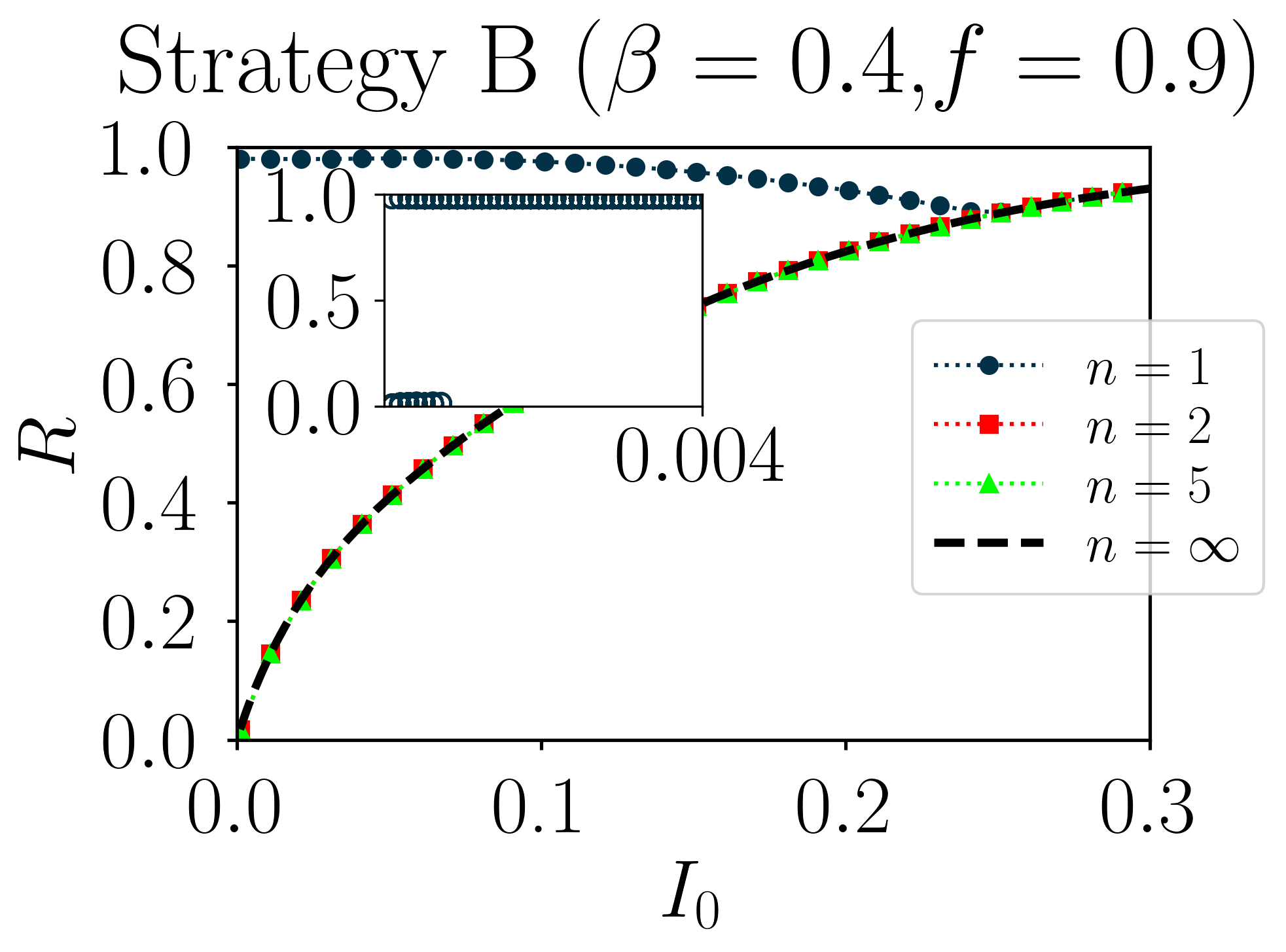}
  \put(30,20){(c)}
\end{overpic}
\begin{overpic}[scale=0.45]{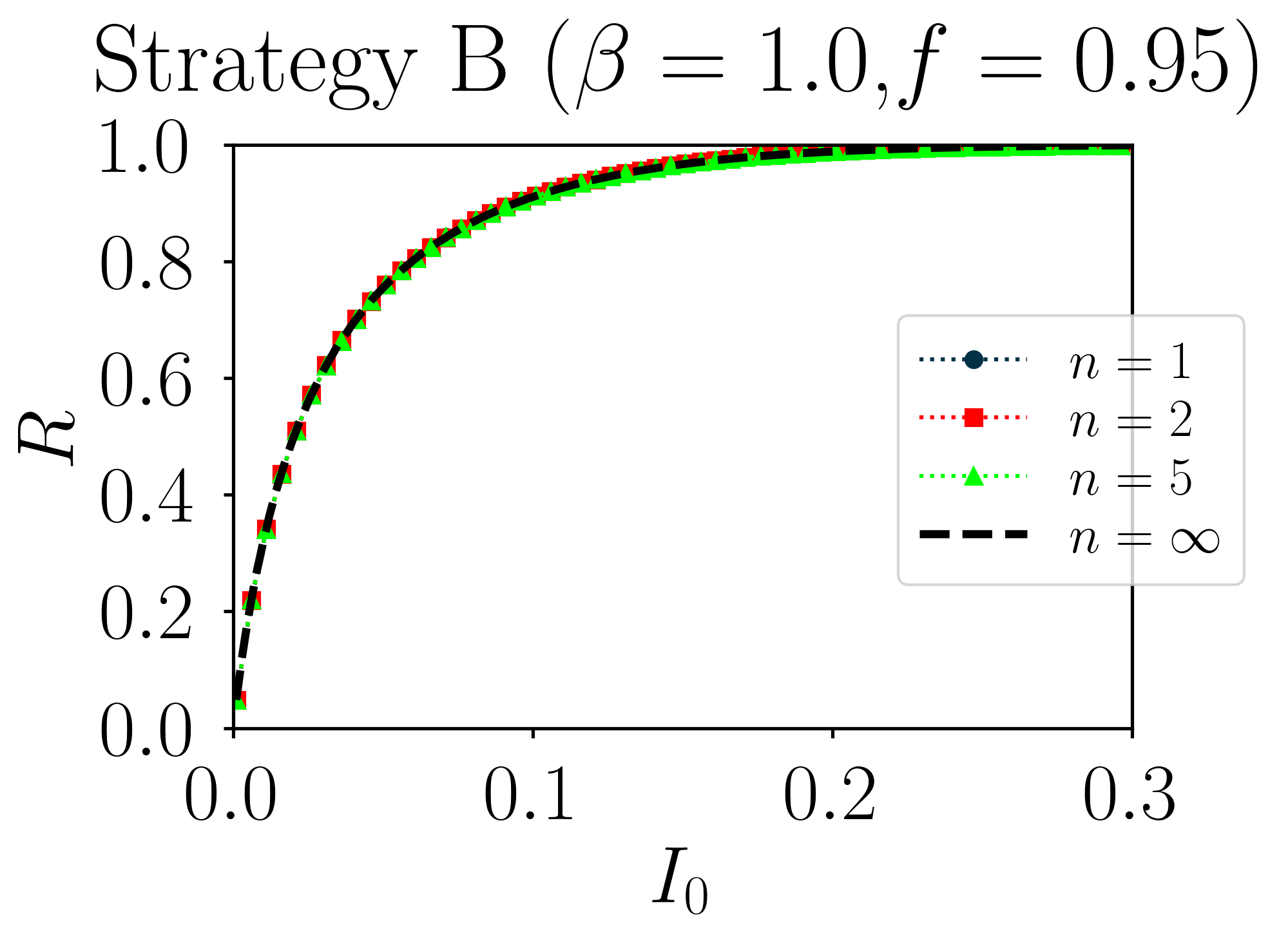}
  \put(30,20){(d)}
\end{overpic}
\vspace{0cm}
\vspace{0.0cm}
\end{center}
\caption{$R$, as a function of
  $I_0$ (symbols) for Strategy A [panels (a)-(b)] and Strategy B [panels (c)-(d)] considering various values of the fatigue threshold $n$, $\beta$ and $f$. The insets show the scatter plots of $R$ vs. $I_0$ for $n=1$ (panels a-c) and $n=2$ (panel b). All simulations were performed on RR networks with $k_C=7$, $k_I=3$ and $N_I=10^6$. Results were
  obtained from $10^2$ stochastic realizations.}\label{fig.I0scatt}
\end{figure}

From Figs.~\ref{fig.I0scatt}a-c, it is clear that $R$ initially increases as a continuous function of $I_0$. However, beyond a critical threshold denoted by $I_0^*$, $R$ exhibits a sharp transition, indicating a backward-type bifurcation because this abrupt change occurs for $R_0<1$. Although this phenomenon was previously observed for the case of $t_b=\infty$ in Ref.~\cite{valdez2023epidemic}, our present results confirm that a backward bifurcation can also occur when isolated individuals reestablish their connections after a time $t_b<\infty$. Notably, when considering the specific parameter values used in Fig.~\ref{fig.I0scatt}b ($\beta=1$ and $f=0.4$), this phenomenon is consistently observed across all the values of $n$. However, Figs.~\ref{fig.I0scatt}a and c, clearly illustrate that a backward bifurcation exclusively emerges for small values of $n$.

Moreover, from Figs.~\ref{fig.I0scatt}a-c, we can see that $R$ displays a non-monotonic behavior as $I_0$ increases.  Following similar reasoning as in Sec.~\ref{sec.ResMicro}, this behavior may arise from the interplay of two opposing mechanisms. On one hand, an increasing number of initially infected people will contribute to a larger cumulative number of recovered individuals at the final stage. However, on the other hand, this increase in $I_0$ also results in a greater number of individuals being isolated which prevents further transmission of the disease.   Consequently, as the second mechanism begins to prevail, we expect a decline in $R$ with increasing $I_0$, as shown in Figs.~\ref{fig.I0scatt}a-c.

Interestingly, for Strategy A, we can see that an additional transition can occur at a critical point denoted as $I_0^{\dag}$, above which the proportion of recovered people $R$ decreases sharply, particularly for small values of $n$ (see Figs. ~\ref{fig.I0scatt}a-b). Therefore, these findings demonstrate that our SIRQ model with fatigue can be highly sensitive to initial conditions when $R_0<1$.

\section{Conclusions}\label{Sec.Conclu}
In this paper, we studied a quarantine model with fatigue on random networks with cliques. Our findings show that quarantine fatigue can undermine the effectiveness of this strategy, leading to an increase in the cumulative number
of cases at the end of an epidemic.

In addition, we investigated the impact of different initial conditions on the nature of the transition point. On one hand, for a single infected patient at the onset of disease transmission, we obtained that an abrupt phase transition can occur at the critical reproduction number  $R_0=1$, particularly in cases where there is little public adherence to quarantine protocols (i.e., small values of $n$). On the other hand, for a macroscopic fraction of initially infected individuals ($I_0$), we found an abrupt transition between a controlled epidemic and a large epidemic for $R_0<1$. Moreover, an increase in $I_0$ was found to lead to one or two abrupt changes in the number of cumulative cases at the end of the epidemic process, especially for small values of $n$.  Therefore, these results suggest that models considering the effect of fatigue should also explore their sensitivity to initial conditions.

As new diseases are expected to emerge in the near future and vaccines will take several months to be developed and approved, it is important to evaluate the viability of applying quarantine measures for extended periods. We hope that our present findings will contribute to the development of new models focused on the relationship between quarantine fatigue and epidemic outbreaks.

\section{Acknowledgements}\label{Sec.Ack}
This work was partially funded by UNMdP  (EXA 956/20) and ANPCyT (PICT 1422/2019 and PICT-2021-I-INVI-00255). %The author thanks XYZ for a critical reading of the manuscript.

\appendix
\counterwithin{figure}{section}
\counterwithin{table}{section}
\renewcommand{\thefigure}{\Alph{section}.\arabic{figure}}
\renewcommand{\thetable}{\Alph{section}.\arabic{table}}

\section{Bipartite Networks}\label{App.Bip}
In this work, we use the configuration model to generate an ensemble of random networks with cliques  (as illustrated in Fig.~\ref{fig.Bip}) that preserve:
1) the total number of nodes or individuals $N_I$, 2) the total number of cliques $N_C$, 3) the probability that a clique contains $k_C$ individuals denoted as $P(k_C)$, and 4) the probability that an individual belongs to $k_I$ cliques denoted as $P(k_I)$. Specifically,  we generate random networks with cliques as bipartite networks by following these steps:
\begin{itemize}
\item Step 1) We generate two sets for the bipartite network: one for cliques with $N_C$ elements and the other for individuals with $N_I$ elements.
\item Step 2) Then we assign $k_C$ stubs or half-links to each clique (sampled from the distribution $P(k_C)$). Similarly, we allocate $k_I$ stubs to each individual (sampled from the distribution $P(k_I)$). Note that these probability distributions must fulfill the condition $\langle k_I \rangle N_I = \langle k_C \rangle N_C$, where $\langle k_I \rangle$ and $\langle k_C \rangle$ denote the mean value of $k_I$ and $k_C$, respectively.
\item Step 3) We randomly select one stub from the list of stubs belonging to cliques and another from the list of stubs belonging to individuals. After that, we create a link by joining these stubs, ensuring that there are no multiple links between individuals and cliques. We repeat this step until one of the lists is empty.
\item Step 4) If there are unconnected stubs, we eliminate them from the bipartite network. Finally, we proceed with a projection process where all individuals connected to the same clique become interconnected with each other.
\end{itemize}
Note that a different algorithm to build random networks with cliques can be found in Ref.~\cite{sobehart2023structural}.

\section{Critical Detection Probability for Different Clique Sizes}\label{App.fckc}

In this section, we investigate the relationship between $f_c$ and $k_C$ in random regular networks with cliques, for a fixed recovery time of $t_r=1$. The findings presented here are valid only for a scenario with a microscopic initial number of infected individuals.

\begin{figure}[H]
\vspace{0.0cm}
\begin{center}
\begin{overpic}[scale=0.45]{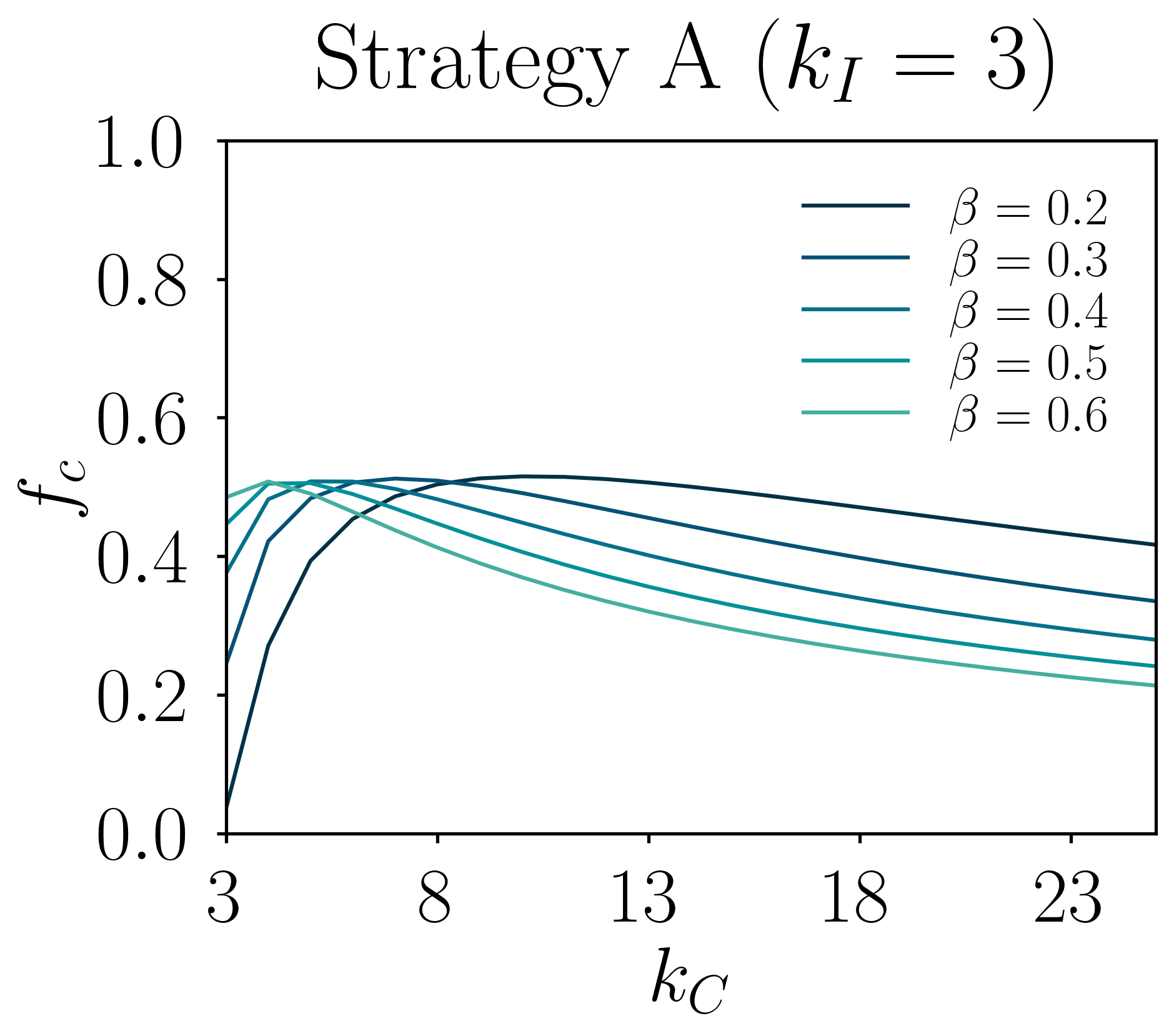}
  \put(80,22){(a)}
\end{overpic}
\vspace{0.0cm}
\begin{overpic}[scale=0.45]{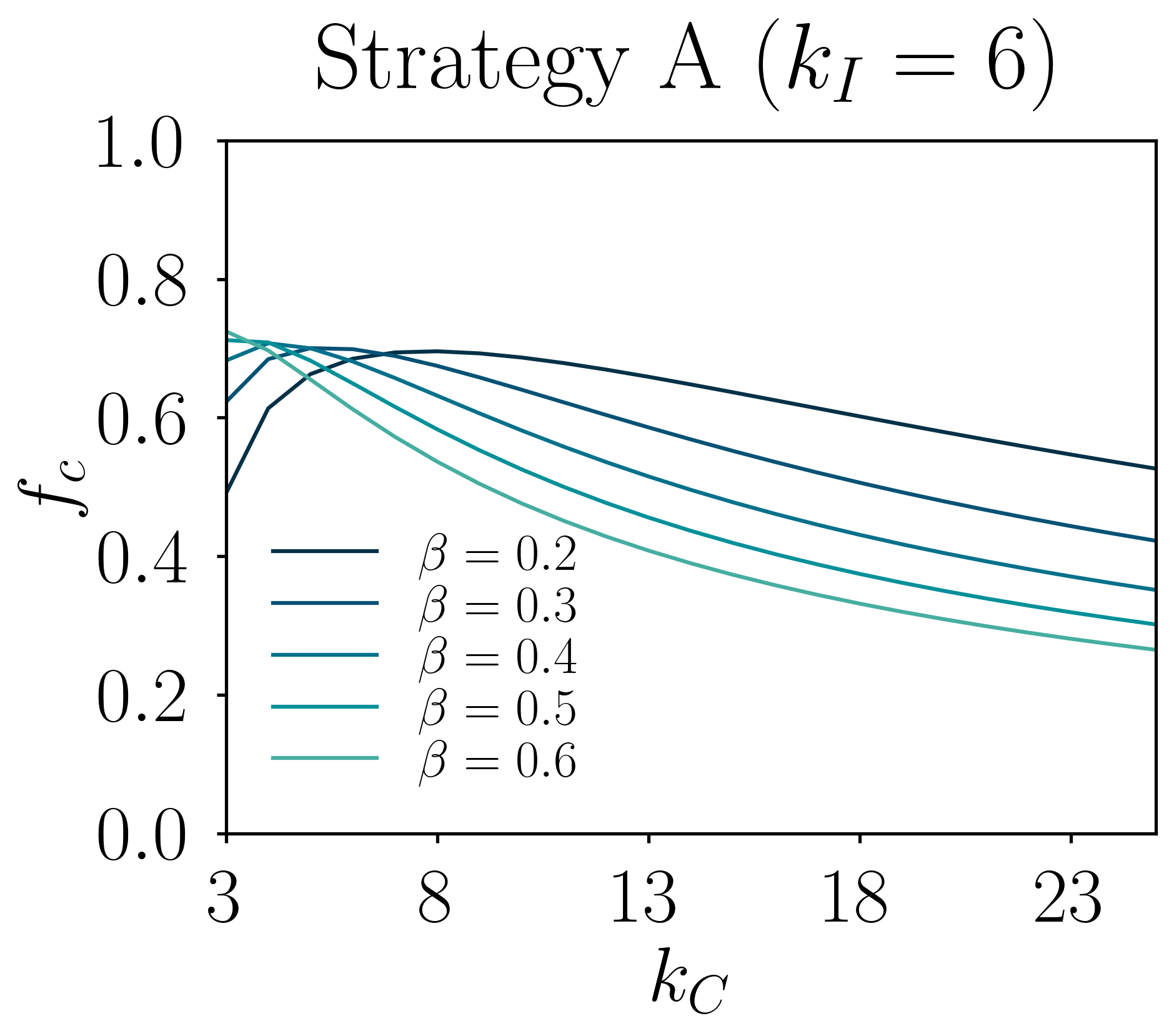}
  \put(80,22){(b)}
\end{overpic}
\vspace{0.0cm}
\begin{overpic}[scale=0.45]{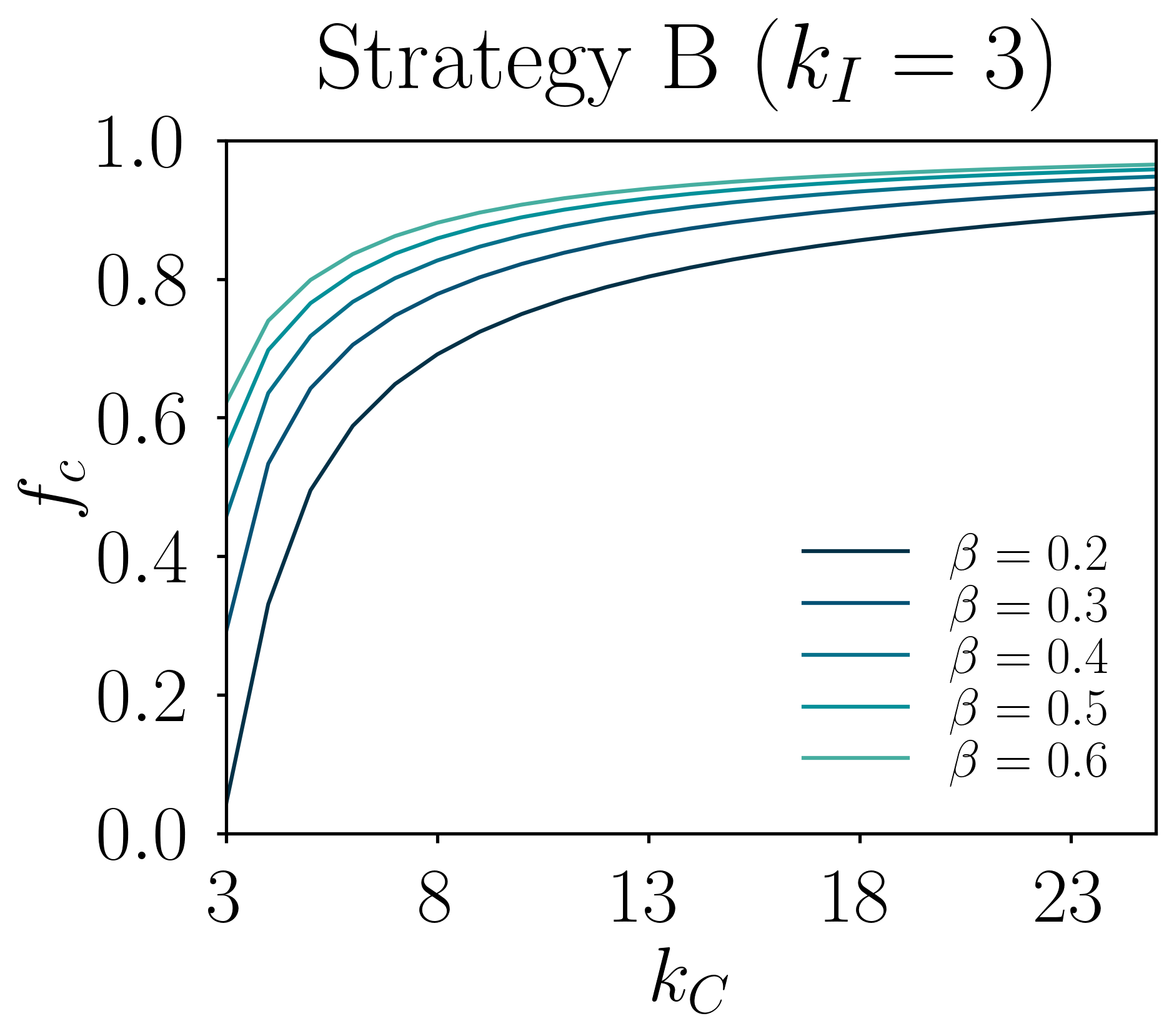}
  \put(30,22){(c)}
\end{overpic}
\begin{overpic}[scale=0.45]{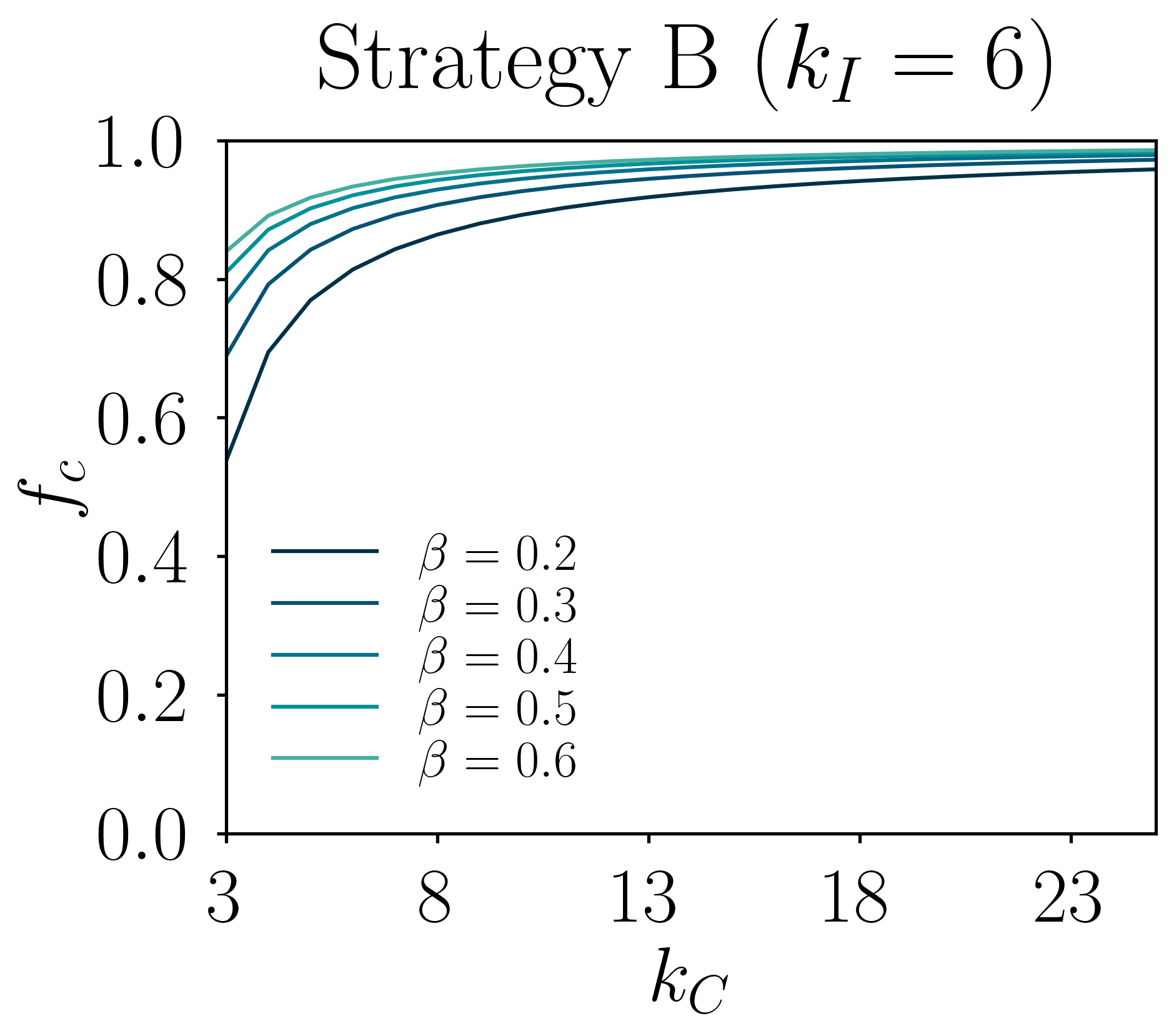}
  \put(80,22){(d)}
\end{overpic}
\vspace{0cm}
\vspace{-0.6cm}
\end{center}
\caption{Critical detection probability $f_c$ as a function of
  $k_C$ for Strategy A [panels (a) and (b)] and Strategy B [panels (c) and (d)] for RR networks with cliques. We compute $f_c$ by solving Eq.~(\ref{eq.StA.R0}) for Strategy A and Eq.~(\ref{eq.StB.R0}) for Strategy B, with $R_0=1$ and considering different values of $\beta$ and $k_I$.}\label{fig.fCvskC}
\end{figure}

In order to compute $f_c$, we solve Eqs.~(\ref{eq.StA.R0}) and (\ref{eq.StB.R0}) for $R_0=1$ and considering several values of $\beta$ and $k_I$. The results for Strategy A (panels a and b) and Strategy B (panels c and d) are shown in Fig.~\ref{fig.fCvskC}.

For Strategy A and large values of $k_C$, we observe that $f_c$ decreases as a function of $k_C$, as expected,  due to two key factors: i)  under Strategy A, individuals are quarantined before they can infect others, and ii) larger cliques are more likely to be placed under quarantine, reducing the likelihood of disease transmission. Consequently, this strategy becomes more effective as clique sizes increase. Interestingly, we also observe the existence of an "optimal" clique size ($k_C^*$) at which $f_c$ reaches its maximum value ($f_c^*$).  Therefore, any detection probability exceeding $f_c^*$ guarantees the prevention of epidemics.

In contrast, for Strategy B, we observe that $f_c$ does not exhibit a local minimum or maximum, and instead grows monotonically with $k_C$ (see panels c and d). In other words,  larger groups require a higher detection probability to prevent the development of an epidemic. This is because under Strategy B, individuals can infect others before being quarantined, allowing the disease to "escape" from the clique even if an infected individual is detected.

\bibliography{bib}
 
\end{document}